\documentclass[11pt, letterpaper, onecolumn, peerreview]{IEEEtran}
\usepackage{graphicx}
\usepackage{amsmath}
\usepackage{amssymb}               
\usepackage{amsfonts}               
\usepackage{amsopn}
\usepackage{xcolor}
\usepackage{colortbl}
\usepackage{multirow}
\usepackage[pdftex]{hyperref}
\usepackage{cite}
\usepackage{stfloats}
\newtheorem{theorem}{\bf Theorem}
\newtheorem{lemma}{\bf Lemma}
\newtheorem{claim}{Claim}
\newtheorem{proposition}{\bf Proposition}

\newtheorem{corollary}{\bf Corollary}
\newtheorem{definition}{\bf Definition}
\newtheorem{example}{Example}

\newtheorem{remark}{\bf Remark}

\def\hY{\hat{Y}}
\def\hy{\hat{y}}

\def\hZ{\hat{Z}}

\def\styp{\mathit{T}_{\epsilon}^n}
\def\stypo{\mathit{T}_{\epsilon'}^n}
\def\stypt{\mathit{T}_{\epsilon''}^n}

\def\be{\begin{equation}}
\def\ee{\end{equation}}
\def\bes{\begin{equation*}}
\def\ees{\end{equation*}}
\def\beq{\begin{eqnarray}}
\def\eeq{\end{eqnarray}}
\def\beqs{\begin{eqnarray*}}
\def\eeqs{\end{eqnarray*}}
\def\ma{{\mathcal A}}

\def\mc{{\mathcal C}}
\def\md{{\mathcal D}}
\def\me{{\mathcal E}}
\def\mf{{\mathcal F}}

\def\ml{{\mathcal L}}
\def\mm{{\mathcal M}}
\def\mn{{\mathcal N}}

\def\mp{{\mathcal P}}
\def\mq{{\mathcal Q}}
\def\mr{{\mathcal R}}
\def\ms{{\mathcal S}}
\def\mt{{\mathcal T}}
\def\mu{{\mathcal U}}
\def\mv{{\mathcal V}}
\def\mw{{\mathcal W}}
\def\mx{{\mathcal X}}
\def\my{{\mathcal Y}}
\def\mz{{\mathcal Z}}

\def\um{\mathbf{m}}

\def\bC{\mathbf{C}}
\def\bU{\mathbf{U}}
\def\F{\mathbf{F}}
\def\bp{\mathbf{P}}

\def\uu{\mathbf{u}}

\def\ux{\mathbf{x}}
\def\uy{\mathbf{y}}
\def\uz{\mathbf{z}}
\def\bX{\mathbf{X}}
\def\bY{\mathbf{Y}}

\def\uhy{\mathbf{\hat{y}}}
\def\bhY{\mathbf{\hat{Y}}}
\def\n{\nonumber\\}
 \def\clap#1{\hbox to 0pt{\hss#1\hss}}
 \def\mathclap{\mathpalette\mathclapinternal}
 \def\mathclapinternal#1#2{%
 \clap{$\mathsurround=0pt#1{#2}$}%
 }
 \def\mathrlap{\mathpalette\mathrlapinternal}
 \def\mathrlapinternal#1#2{%
 \rlap{$\mathsurround=0pt#1{#2}$}
 }

\begin{document}

\author{Mohammad~Hossein~Yassaee,~\IEEEmembership{Student Member,~IEEE,} and ~Mohammad~Reza~Aref
\thanks{ This work was supported by Iranian-NSF under grant No. 88114.46-2010. The material in this paper was
presented in part at the IEEE International Symposium on Information Theory,
Toronto, Canada, July 2008 and the IEEE International Symposium on Information Theory, Seoul, Korea, June 2009.}
\thanks{The authors are with the Information Systems and Security Lab (ISSL), Department of Electrical Engineering, Sharif University of Technology, Tehran, Iran (e-mail: yassaee@ee.sharif.edu; aref@sharif.edu).}
}
\title{Slepian-Wolf Coding Over Cooperative Relay Networks}
\maketitle
\begin{abstract}
This paper deals with the problem of multicasting a set of discrete memoryless correlated  sources (DMCS) over a cooperative relay network. Necessary conditions with cut-set interpretation are presented. A \emph{Joint source-Wyner-Ziv encoding/sliding window decoding} scheme is proposed, in which decoding at each receiver is done with respect to an ordered partition of other nodes. For each ordered partition a set of feasibility constraints is derived. Then, utilizing the sub-modular property of the entropy function and a novel geometrical approach, the results of different ordered partitions are consolidated, which lead to sufficient conditions for our problem. The proposed scheme achieves operational separation between source coding and channel coding. It is shown that sufficient conditions are indeed necessary conditions in two special cooperative networks, namely, Aref network and finite-field deterministic network. Also, in Gaussian cooperative networks, it is shown that reliable transmission of all DMCS whose Slepian-Wolf region intersects the cut-set bound region within a constant number of bits, is feasible. In particular, all results of the paper are specialized to obtain an achievable rate region for cooperative relay networks which includes relay networks and two-way relay networks. 
\end{abstract}
\section{Introduction}
Consider a group of $K$ sensors measuring a common phenomenon, like weather. In this paper, we investigate a communication scenario in which some sensors desire to obtain measurements of the other nodes with the help of some existing relay nodes in the network. In the language of information theory, we can consider  measurements of sensors as outputs of discrete memoryless correlated sources and model the communication network as a cooperative relay network in which each node can simultaneously be a transmitter, a relay and a receiver. So the problem can be defined as below:\par
\emph{Given a set of sources $U_{\ma}=\{U_{a_j}:a_j\in\ma\}$ observed at nodes $\ma=\{a_1,\cdots,a_M\}\subseteq\mv$ respectively ($\mv=\{1,\cdots,V\}$ is the set of nodes in the network) and a set of receivers at nodes $\md=\{d_1,\cdots,d_K\}\subseteq\mv$ which is not necessarily disjoint from $\ma$, what conditions must be satisfied to enable us to reliably multicast $U_{\ma}$ to all the nodes in $\md$ over the cooperative relay network?} \par 
The problem of Slepian-Wolf (SW) coding over multi-user channels has been considered for some special networks. First in \cite{tuncel}, Tuncel investigated the problem of multicasting a source over a broadcast channel with side information at the receivers. He proposed a joint source-channel coding scheme which achieves \emph{operational separation} between source coding and channel coding in the sense that the source and channel variables are separated. He also proved the optimality of his scheme. In a recent work \cite{gunduz}, this problem was generalized to the problem of lossy multicasting of a source over a broadcast channel with side information. In \cite{babu}, a necessary and sufficient condition for multicasting a set of correlated sources over acyclic Aref networks \cite{aref} was derived. The problem of multicasting correlated sources over networks was also studied in the network coding literature \cite{ho,effros}.\par
Cooperative relay network has been widely studied in terms of achievable rate region for relay networks \cite{xie,kramer2005}, multiple access relay channels \cite{marc} and multi-source, multi-relay and multi-destination networks \cite{xie:2007}. In all the mentioned works, two main strategies of Cover and El Gamal for relay channels \cite{cover}, namely, \emph{decode and forward} (DF) and \emph{compress and forward} (CF) were generalized for the cooperative relay networks. In a more general setting \cite{goldsmith}, G\"{u}nd\"{u}z, et.al., consider a compound multiple access channel with a relay in which three transmitters where, one of them acts as a relay for the others, want to multicast their messages to the two receivers. Several Inner bounds to the capacity region of this network were derived using DF, CF and also structured lattice codes. Although finding the capacity of the simple relay channel is a longstanding open problem, an approximation for the Gaussian relay network with multicast demands has been recently found in \cite{aves:isit,aves:phd,aves:sub}. In these works, the authors propose a scheme that uses the Wyner-Ziv coding at the relays and a distinguishability argument at the receivers.\par
 In this paper, we first study the problem of multi-layer Slepian-Wolf coding of multi-component correlated sources, in which each source should encode its components according to a given hierarchy. Using the sub-modularity of the entropy function and a covering lemma, we prove an identity which states that for any points of SW-region with respect to joint encoding/decoding of the components, there exists a multi-layer SW-coding which achieves it. To the best of our knowledge, this identity is new and we call it the SW-identity. Then, we propose a \emph{joint Source-Wyner-Ziv encoding/sliding window decoding} scheme for Slepian-Wolf coding over cooperative networks. In this scheme, each node compresses its channel observation using Wyner-Ziv coding and then jointly maps its source observation and compressed channel observation to a channel codeword. For decoding, each receiver uses sliding window decoding with respect to an ordered partition of other nodes. For each ordered partition, we obtain a set of DMCS which can reliably be multicast over the cooperative relay network. By utilizing the SW-identity, we obtain the union of the sets of all feasible DMCS with respect to all ordered partitions. Our scheme results in \emph{operational separation} between the source and channel coding. In addition, this scheme does not depend on the graph of the network, so the result can easily be applied to any arbitrary network. 
  We show that the sufficient conditions for our scheme, are indeed necessary conditions for the Slepian-Wolf coding over arbitrary Aref networks and linear finite-field cooperative relay networks. Moreover, we prove the feasibility of multicasting of all DMCS whose Slepian-Wolf region overlap the cut-set bound within a constant number of bits over a Gaussian cooperative relay network. This establishes a large set of DMCS that belongs to the set of DMCS which can reliably be multicast in the operational separation sense.  Note that the model considered in this paper, encompasses the model of multiple access channel with correlated sources. So the set of feasible DMCS in the operational separation sense is a subset of all feasible DMCS. We extract an achievable rate region for cooperative relay networks by reducing sufficient conditions for reliable multicasting. We show that this achievable rate region subsumes some recent achievable rates based on the CF strategy \cite{kramer2005,yassaee}. In addition, we estimate the capacity region of Gaussian cooperative relay networks within a constant number of bits from the cut-set bound. Our result improves capacity approximation of Gaussian relay networks given in \cite{aves:sub}.\par
The rest of the paper is organized as follows. In section \ref{sec:2}, we introduce notations and definitions used in this paper. Section \ref{sec:3} derives necessary conditions for reliable multicasting of DMCS over cooperative networks. Section \ref{sec:4} studies the multi-layer Slepian-Wolf coding, in particular, a novel identity related to the entropy function is derived. In section \ref{sec:5}, we obtain feasibility constraints which are the main results of the paper. In sections \ref{sec:6} and \ref{sec:7}, we derive necessary and sufficient conditions for multicasting of DMCS over some classes of semi-deterministic networks and Gaussian cooperative relay networks, respectively. Section \ref{sec:8} employs results of the previous sections to derive an inner bound and an outer bound for the capacity region of a cooperative relay networks. Section \ref{sec:9} concludes the paper.                 
\section{Preliminaries and Definitions}\label{sec:2}
\subsection{Notation}
 We denote discrete random variables with capital letters, e.g., $X$, $Y$, and their realizations with lower case letters $x$, $y$. A random variable $X$ takes values in a set $\mx$. We use $|\mx|$ to denote the cardinality of a finite discrete set $\mx$, and $p_X(x)$ to denote the probability mass function (p.m.f.) of $X$ on $\mx$, for brevity we may omit the subscript $X$ when it is obvious from the context. We denote vectors with boldface letters, e.g. $\ux$, $\uy$. The superscript identifies the number of samples to be included in a given vector, e.g., $X^i=(X_1,\cdots,X_i)$. We use $\styp(X)$ to denote the set of $\epsilon$-strongly typical sequences of length $n$, with respect to p.m.f. $p_X(x)$ on $\mx$. Further, we use $\styp(Y|\ux)$ to denote the set of all $n$-sequences $\uy$ such that $(\ux,\uy)$ are jointly typical, w.r.t. $p_{XY}(x,y)$. We denote the vectors in the $j$th block by a subscript $[j]$. For a given set $\ms$, we use the shortcuts $X_{\ms}=\{X_i:i\in\ms\}$ and $R_{\ms}=\sum_{i\in\ms}R_i$. We use $\ms\backslash\mt$ to denote the set theoretic difference of $\ms$ and $\mt$. We say that $a_n\stackrel{.}{\le}2^{nb}$, if for each $\epsilon>0$ and sufficiently large $n$, the relation $a_n\le 2^{n(b-\epsilon)}$ holds.
\subsection{Sub-modular Function}
Let $\mv$ be a finite set and $2^{\mv}$ be a power set of it, i.e., the collection of all subsets of $\mv$. A function $f:2^{\mv}\rightarrow\mathbb{R}$ is called sub-modular, if for each $\ms,\mt\subseteq\mv$,
\be
f(\ms\cap\mt)+f(\ms\cup\mt)\le f(\ms)+f(\mt)
\ee 
Function $f$ is called super-modular, if $-f$ is sub-modular. Given two sets $\ms,\mt$ and a sub-modular function $f$, we define $f(\ms|\mt)\triangleq f(\ms\cup\mt)-f(\mt)$.\\
Let $X_{\ma}$ be DMCS with distribution $p(x_{\ma})$. For each $\ms\subseteq\ma$, we define the entropy function $h$ as $h(\ms)=H(X_{\ms})$ where $H(X)$ denotes the entropy of random variable $X$. It is well-known that the entropy function $h$ is a sub-modular function over the set $\ma$ \cite{yeoung:book}. The sub-modularity property of the entropy function plays an essential role in the remainder of the paper, (in contrast to the non-decreasing property of the entropy, i.e, $h(\ms)\ge h(\mt),\ \forall \mt\subseteq\ms$).     
\subsection{Some Geometry}
A polytope is a generalization of polygon to a higher dimension. Point, segment and polygon are polytopes of dimension $0$, $1$ and $2$, respectively. A polytope of dimension $d\ge 3$ can be considered as a space bounded by a set of polytopes of dimension $d-1$. The boundary polytope of dimension $d-1$ is called facet. For a given polytope $\bp$, a collection of polytopes $\{\bp_1,\cdots,\bp_n\}$ is called a \emph{closed covering} of $\bp$, if $\bp=\cup_{i=1}^n \bp_i$. 
\begin{lemma}\label{le:cover}
Let $\bp$ be a polytope and $\mf=\{\bp_1,\bp_2,\cdots,\bp_n\}$ be a collection of polytopes with the same dimension as $\bp$. If $\bp$ and $\mf$ satisfy the following conditions:
\begin{enumerate}
\item $\forall i:\quad\bp_i\subset\bp$
\item Each facet of $\bp$ is covered by some facets of some polytopes $(\bp_{i1},\cdots,\bp_{ik})$.
\item For each facet of $\bp_i$ inside $\bp$, there is $\bp_j\neq\bp_i$ such that 
$\bp_i$ and $\bp_j$ have only that facet as the common part.
\end{enumerate}
\par then $\mf$ is a \emph{closed covering} of $\bp$. 
\end{lemma}
\begin{proof}
The proof is provided in the Appendix \ref{app:1}.
\end{proof}
Lemma \ref{le:cover} provides a powerful tool for dealing with the regions which are described with a set of inequalities. 
\begin{definition}
\label{def:majorize}
A point $Q=(q_1,\cdots,q_d)$
in $\mathbb{R}^d$ is said to majorize point $P =(p_1,\cdots,p_d)$, if $q_i\ge p_i$ for all $i$. In addition, point $Q$ is said to majorize  set $\mp$ (denoted by $Q\succ\mp$), if there exists a point $X\in\mp$ which is majorized by $Q$.
\end{definition}
It is easy to show that majorization has the following simple property:
  \be
     \label{eq:maj-property}
     Q\succ \mp_1\cup\mp_2 \quad \Leftrightarrow\quad   Q\succ \mp_1\ \mbox{or}\ Q\succ \mp_2 
  \ee     
\begin{definition}
\label{def:associate}
Let $f$ be a sub-modular function over the set $\mv$. The \emph{essential polytope} associated with $f$ is:
\be
\label{eq:esspoly}
\bp_f=\{\ux\in\mathbb{R}^{|\mv|}: x_{\mv}= f(\mv)\ \mbox{and}\ \forall \ms\subset\mv, x_{\ms}\ge f(\ms|\ms^C)\}
\ee
where $\ux=[x_1,x_2,\cdots,x_{|\mv|}]$ and $x_{\ms}=\sum_{i\in\ms}x_i$.
\end{definition} 
\par The essential polytope of the sub-modular function $f$ over the set $\mv$ is a polytope of dimension $|\mv|-1$, which has $2^{|\mv|}-2$ facets, each corresponding to intersection of hyperplane $x_{\mt}=f(\mt|\mt^C)$ with $\bp_{f}$ for each non-empty subset $\mt\subset\mv$. By $\F_{f,\mt}$, we denote  the facet corresponding to the subset $\mt$. Since $g(\mt)=f(\mt|\mt^C)=f(\mv)-f(\mt)$ is a super-modular function, one can easily show that $\F_{f,\mt}$ is a non-empty polytope of dimension $|\mv|-2$ (see for example, \cite{polytope}) .
\begin{lemma}
\label{le:facet}
The facet $\F_{f,\mt}$ of polytope $\bp_f$ can be decomposed to projections of $\bp_f$ on $\mathbb{R}^{\mt}$ and $\mathbb{R}^{\mt^C}$ (in which $\mathbb{R}^{\ms}$ stands for the space $\{\ux\in\mathbb{R}^{|\mv|}:\forall s\in\ms^C, x_{s}=0\}$). More precisely, 
\be
\label{eq:app}
\F_{f,\mt}=\{\ux\in\mathbb{R}^{|\mv|}:\ux_{\mt}\in \F_{f,\mt}^{(1)}, \ux_{\mt^C}\in\F_{f,\mt}^{(2)}\}
\ee
where 
\be
\label{eq:pface1}
\F_{f,\mt}^{(1)}=\{\ux\in\mathbb{R}^{\mt}: x_{\mt}=f(\mt|\mt^C),\ \mbox{and}\ \forall\ms\subset\mt, x_{\ms}\ge f(\ms|\ms^C)\}
\ee
and
\be
\label{eq:pface2}
\F_{f,\mt}^{(2)}=\{\ux\in\mathbb{R}^{\mt^C}: x_{\mt^C}=f(\mt^C),\ \mbox{and}\ \forall\ms\subset\mt^C, x_{\ms}\ge f(\ms|\mt^C\backslash\ms)\}.
\ee
Moreover, $\F_{f,\mt}^{(1)}$ and  $\F_{f,\mt}^{(2)}$ are the essential polytopes of the functions $f_1:2^{\mt}\rightarrow\mathbb{R}$ and $f_2:2^{\mt^C}\rightarrow\mathbb{R}$ respectively, where $f_1(\ms)=f(\ms|\mt^C)$ and $f_2(\ms)=f(\ms)$.
\end{lemma} 
\begin{proof}
The proof is provided in Appendix \ref{app:2}.
\end{proof}
%
\begin{lemma}[\cite{polytope}]
\label{le:sum-polytope}
Let $f_1$ and $f_2$ be two sub-modular functions defined on a set $\mv$. Then,
\be
\bp_{f_1+f_2}=\bp_{f_1}+\bp_{f_2}
\ee
where the sum of  two sets is defined as $\mx+\my=\{x+y:x\in\mx,y\in\my\}$. 
\end{lemma} 
\subsection{System Model}
A cooperative relay network is a discrete memoryless network with $V$ nodes $\mv =\{1, 2,\cdots, V\}$, and a channel of the form
\[
(\mx_1,\mx_2,\cdots,\mx_V,p(y_1,y_2,\cdots,y_V|x_1,x_2,\cdots,x_V),\my_1,\my_2,\cdots,\my_V).
\]
At each time $t = 1, 2,\cdots$, every node $v\in\mv$ sends an input $X_{v,t}\in\mx_v$, and receives an output $Y_{v,t}\in\my_v$, which are related via $p(Y_{1,t},\cdots , Y_{V,t}|X_{1,t}, . . . ,X_{V,t})$.
\begin{definition}[Reliable multicasting of correlated sources over cooperative networks]
\label{def:sw}
Let $\ma$ and $\md$ be two subsets of $\mv$ corresponding to the set of the sources and the destinations, respectively. We say that the set of DMCS, $U_\ma$, can reliably be multicast over discrete memoryless cooperative network, to all nodes in $\md$, if there exists a sequence of a pair of positive integers $(s_n,r_n)$ such that $s_n\rightarrow\infty,\ r_n\rightarrow\infty,\ \dfrac{r_n}{s_n}\rightarrow 1$ as $n\rightarrow\infty$ and a sequence of encoding functions 
\[ f_{v,t}^{(s_n)}:\mu_v^{s_n}\times\my_v^{t-1}\rightarrow\mx_v\quad \mbox{for}\quad t=1,\cdots,r_n\] 
at all nodes $v\in\mv$, where, for the non-source nodes we let $\mu_v=\emptyset$ 
and a set of decoding functions defined at each node $d_i\in\md$;
\[g_{d_i}^{(s_n,r_n)}:\mu_{d_i}^{s_n}\times\my_{d_i}^{r_n}\rightarrow\mu_{\ma}^{s_n}\]
such that the probability of error
\[
P_{e,d_i}^{(s_n,r_n)}=\Pr\left(g_{d_i}^{(s_n,r_n)}(U_{d_i}^{s_n},Y_{d_i}^{r_n})\neq U_{\ma}^{s_n}\right)
\] 
vanishes for all $d_i\in\md$ as $n$ goes to the infinity.
 \end{definition}
According to Definition \ref{def:sw}, the joint probability distribution of the random variables factors as,
\be
p(\uu_{\ma},\ux_{\mv},\uy_{\mv})=\prod_{j=1}^{s_n} p(u_{\ma,j})\prod_{t=1}^{r_n}\prod_{v=1}^V p(x_{v,t}|y_v^{t-1},\uu_v)p(y_{\mv,t}|x_{\mv,t})
\ee 
\begin{remark}
The network model described in the Definition \ref{def:sw} includes several network models such as MAC with feedback, relay networks and multi-way channels (i.e., a generalization of the two-way channel). 
\end{remark}
\section{Cut-set type necessary conditions for reliable multicasting}\label{sec:3}
In this section, we prove necessary conditions for reliable multicasting of correlated sources over cooperative network. 
\begin{proposition}
 \label{pro:ob}
 A set of DMCS $U_{\ma}$ can reliably be multicast over a cooperative network, only if there exists a joint p.m.f. $p(x_{\mv})$ such that 
  \begin{equation}
  \label{eq:sw2}
 H(U_{\ms}|U_{\ma\backslash\ms})< \min_{d_i\in\md\backslash\ms}\min_{\mv\supseteq\mw\supseteq\ms:\atop d_i\in\mw^C} I(X_{\mw};Y_{\mw^C}|X_{\mw^C})
 \end{equation}
 \end{proposition}
\begin{proof}
 Using Fano's inequality, imposing the condition $P_{e,d_i}^{(s_n,r_n)}\rightarrow 0$ as $n\rightarrow\infty$, it follows that:
  \begin{equation}
 \label{eq:ob}\forall \ms\subseteq\mv, d_i\in\md\backslash\ms: \frac{1}{s_n}H(U_{\ma}^{s_n}|Y_{d_i}^{r_n},U_{d_i}^{s_n})\leq\epsilon_n
 \end{equation}
  with $\epsilon_n\rightarrow 0$ as $n\rightarrow\infty$. We also have $\frac{1}{s_n}H(U_{\ms}^{s_n}|U_{\ma\backslash\ms}^{s_n}Y_{d_i}^{r_n}U_{d_i}^{s_n})\leq\epsilon_n$. For each $(\mw,d_i)$ such that $\ms\subseteq\mw\subseteq\mv$ and $d_i\in\mw^C$, we have:
  \begin{align}
 H(U_{\ms}|U_{\ma\backslash\ms})&=\frac{1}{s_n}H(U_{\ms}^{s_n}|U_{\ma\backslash\ms}^{s_n})
\label{fa:1}\\ &=\frac{1}{s_n}(I(U_{\ms}^{s_n};Y_{d_i}^{r_n}|U_{\ma\backslash\ms}^{s_n})+H(U_{\ms}^{s_n}|U_{\ma\backslash\ms}^{s_n}Y_{d_i}^{r_n}))
 \label{fa:2}\\
 &\leq\frac{1}{s_n}I(U_{\ms}^{s_n};Y_{\mw^C}^{r_n}|U_{\ma\backslash\ms}^{s_n})+\epsilon_n
 \label{fa:3}\\
&=\frac{1}{s_n}\sum_{i=1}^{r_n}
I(U_{\ms}^{s_n};Y_{\mw^C,i}|U_{\ma\backslash\ms}^{s_n}Y_{\mw^C}^{i-1}X_{\mw^C,i})+\epsilon_n
\label{fa:4}\\
 &=\frac{1}{s_n}\sum_{i=1}^{r_n}H(Y_{\mw^C,i}|U_{\ma\backslash\ms}^{s_n}Y_{\mw^C}^{i-1}X_{\mw^C,i})-
 H(Y_{\mw^C,i}|U_{\ma}^{s_n}Y_{\mw^C}^{i-1}X_{\mw^C,i})+\epsilon_n
 \label{fa:5}\\
 &\leq\frac{1}{s_n}\sum_{i=1}^{r_n}H(Y_{\mw^C,i}|X_{\mw^C,i})-H(Y_{\mw^C,i}|U_{\ma}^{s_n}Y_{\mv}^{i-1}X_{\mv,i})+\epsilon_n
\label{fa:6}\\
&=\frac{1}{s_n}\sum_{i=1}^{r_n}I(X_{\mw,i};Y_{\mw^C,i}|X_{\mw^C,i})+\epsilon_n
\label{fa:7}\\
 &=\frac{r_n}{s_n} I(X_{\mw,Q};Y_{\mw^C,Q}|X_{\mw^C,Q},Q)+\epsilon_n
\label{fa:8}\\ 
 &\leq\frac{r_n}{s_n}I(X_{\mw,Q};Y_{\mw^C,Q}|X_{\mw^C,Q})+\epsilon_n
\label{fa:9}\\
 &\rightarrow I(X_{\mw};Y_{\mw^C}|X_{\mw^C})
\label{fa:10}
 \end{align}
\noindent where \eqref{fa:4} follows from the fact that $X_{\mw^C,i}$ is a function of $(Y_{\mw^C}^{i-1},U_{\mw^C\cap\ma}^{s_n})$ and the fact that $\mw^C\cap\ma\subseteq\ma\backslash\ms$, \eqref{fa:6} follows since conditioning reduces entropy, \eqref{fa:7} follows because $(U_{\ma}^{s_n},Y_{\mv}^{i-1})-X_{\mv,i}-Y_{\mv,i}$ form a Markov chain, \eqref{fa:8} is obtained by introducing a time-sharing random variable $Q$ which is uniformly distributed over the set $\{1,2,\cdots,r_n\}$ and is independent of everything else, \eqref{fa:10} follows by allowing $s_n,r_n\rightarrow\infty$ with $\frac{r_n}{s_n}\rightarrow 1$ and defining $Y_{\mv}\triangleq Y_{\mv,Q}$ and $X_{\mv}\triangleq X_{\mv,Q}$.
\end{proof}
\section{Multi-Layer Slepian-Wolf Coding}\label{sec:4}
Before describing our scheme and the related results, in this section, we deal with the problem of \emph{multi-layer Slepian-Wolf coding} (ML-SW). Study of the ML-SW enables us to find a new tool to analyze the main problem. In the previous works (for example \cite{mlsw}, \cite{mlsw1}), ML-SW is used to describe a source with some small components (for example, by a binary representation of it) and then successively encoding these components with SW-coding instead of encoding the whole source at once. For example, if we describe an i.i.d. source $S$ by $(X,Y)$, i.e., $S=(X,Y)$, instead of encoding $S$ by $R=H(S)$ bits/symbol, we can first describe $X$ by $R_X=H(X)$ bits/symbol and then apply SW-coding to describe $Y$ by $R_Y=H(Y|X)$ bits/symbol, assuming that the receiver knows $X$ from decoding the previous layer information as a side information. Since the total bits required to describe $S$ in two layers is $R_X+R_Y=H(X,Y)=H(S)$, it follows that there is no loss in the two-layer SW-coding compared with the jointly encoding of the source components. A natural question is: \emph{How can this result be generalized to a more general setting of multi-terminal SW-coding}?
\begin{figure}[t]
	\centering
		\includegraphics[scale=.8]{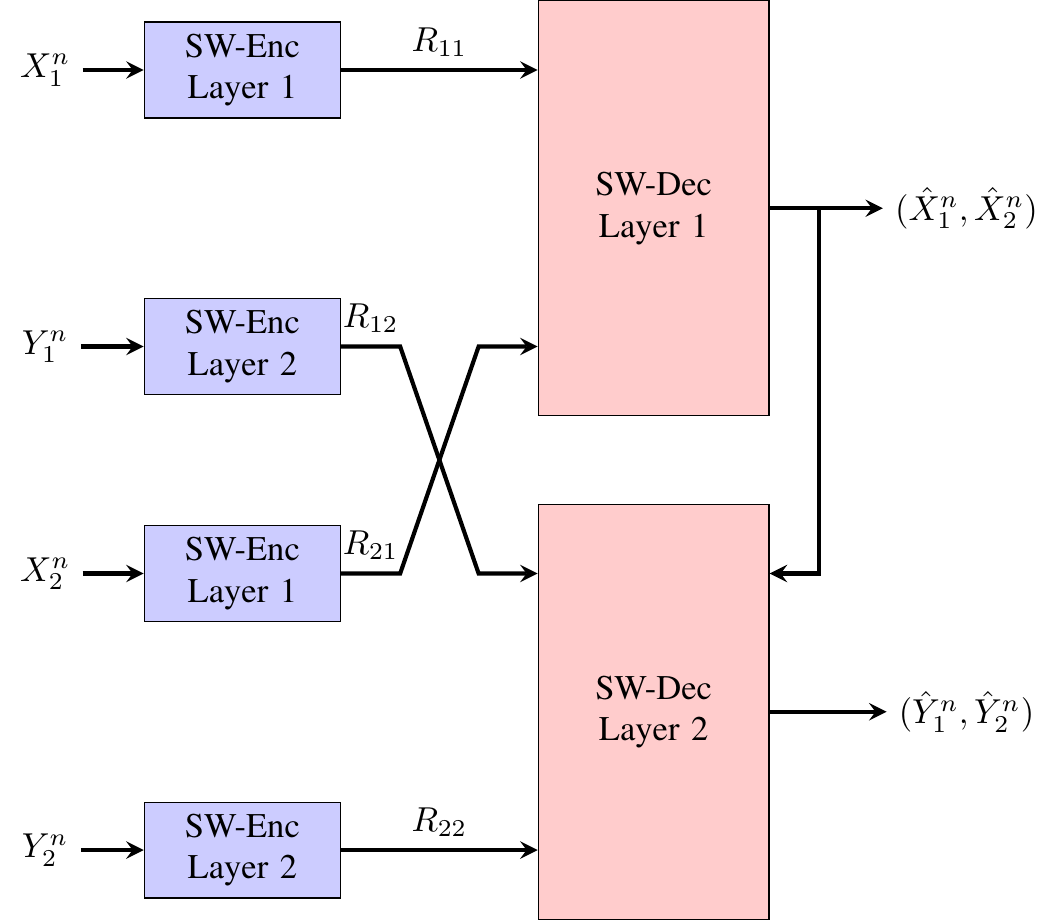}
	\caption{Two-Layer Slepian-Wolf coding for a pair of two-component correlated sources. This coding is suboptimal in the sense that it does not achieve the entire of Slepian-Wolf coding.  }
	\label{fig:2L}
\end{figure}
At first, let us look at the two-terminal SW-coding. Suppose two sources $S_1=(X_1,Y_1)$
and $S_2=(X_2,Y_2)$ are given. Joint SW-coding yields that lossless description of $(S_1,S_2)$ with rates $(R_1,R_2)$ is feasible, provided that $(R_1,R_2)\in\{(r_1,r_2):r_1\ge H(X_1Y_1|X_2Y_2),r_2\ge H(X_2Y_2|X_1Y_1), r_1+r_2\ge H(X_1X_2Y_1Y_2)\}$. Now suppose the following simple ML-SW. Assume in the first layer, $X_1$ and $X_2$ are encoded by SW-coding with rates $(R_{11},R_{21})$ and in the next layer $Y_1$ and $Y_2$ are encoded by SW-coding with rates $(R_{12},R_{22})$ assuming that the receiver knows $(X_1,X_2)$ from decoding of the previous layer information (See Fig. \ref{fig:2L}). The lossless description of $(S_1,S_2)$ in this manner is possible, if:
\begin{align*}
R_1=R_{11}+R_{12}&\ge H(X_1|X_2)+H(Y_1|X_1X_2Y_2)\ge H(X_1Y_1|X_2Y_2)\\
R_2=R_{21}+R_{22}&\ge H(X_2|X_1)+H(Y_2|X_1X_2Y_1)\ge H(X_2Y_2|X_1Y_1)\\        
R_1+R_2&\ge H(X_1X_2)+H(Y_1Y_2|X_1X_2)=H(X_1X_2Y_1Y_2)
\end{align*}
\begin{figure}[t]
	\centering
		\includegraphics[scale=.7]{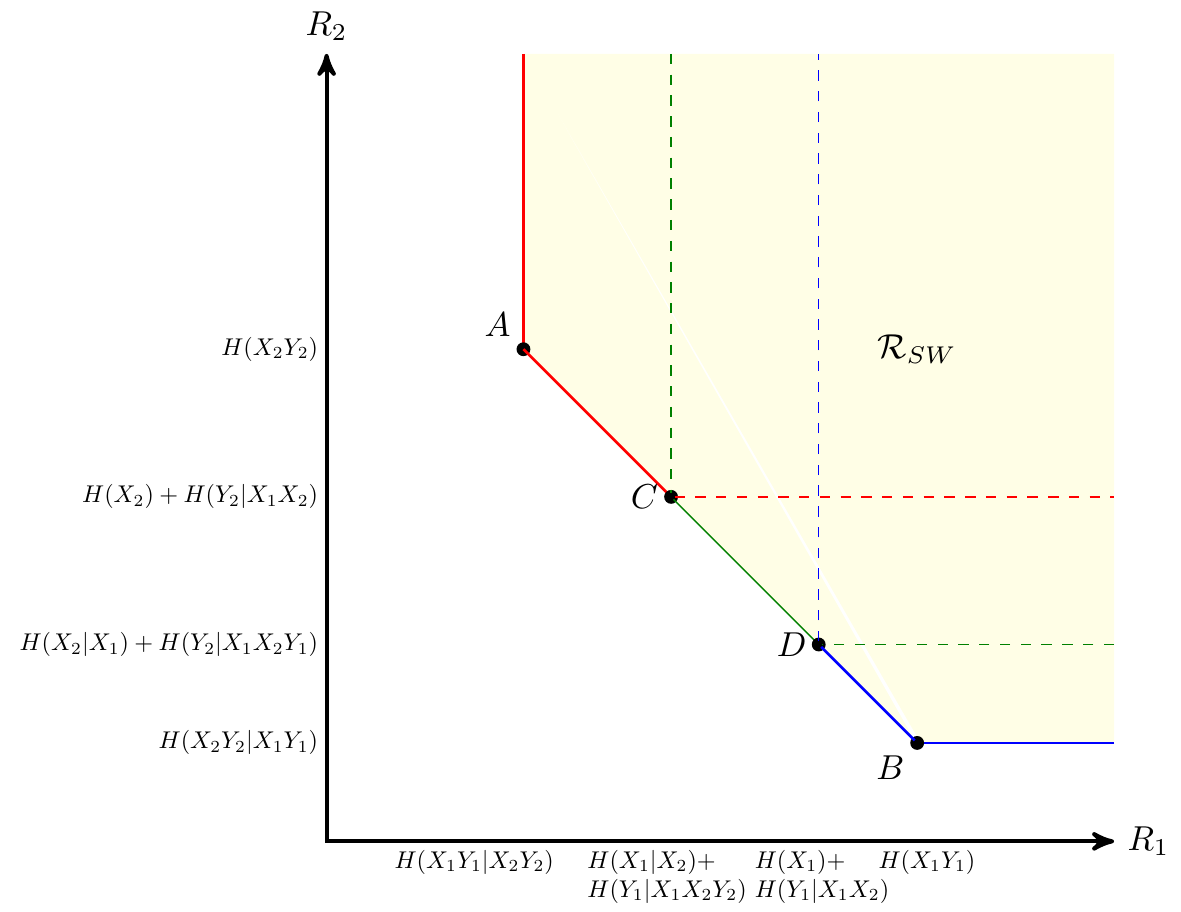}
	\caption{Slepian-Wolf rate region vs rate regions with two and three layers Slepian-Wolf coding. Segments $AC$ correspond to the three-layer SW-coding, in which in the first layer, $X_2$ is encoded, then in the second layer $(Y_2,X_1)$ is encoded assuming that $X_2$ is already decoded at the receiver and in the third layer $Y_1$ is encoded assuming that $(X_2,Y_2,X_1)$ is already available at the receiver. Segment $CD$ corresponds to two-layer SW-coding of the Fig. \ref{fig:2L}. Segment $DB$ is obtained from a similar three layer SW-coding to that of segment $AC$. Notice that each corner point of any multi-layer SW-coding that lies inside the SW-region is coincident to a corner point of another multi-layer SW-coding.    }
	\label{fig:SW}
\end{figure}
This shows that this simple layering can not achieve all the points in the SW-region, in particular the corner points $A=(H(X_1Y_1|X_2Y_2),H(X_2Y_2))$ and $B=(H(X_1Y_1),H(X_2Y_2|X_1Y_1))$ can not be achieved by this scheme(See Fig. \ref{fig:SW}). But the point $A$ can be achieved by successive SW-coding of $X_2$, $Y_2$, $X_1$ and $Y_1$ regarding that the previous sources are available at the receiver. This method suggests that instead of dividing the SW-coding in two layers, SW-coding can be performed in three layers: in the first layer $X_2$ is described for the receiver with rate $R_{21}\ge H(X_2)$, in the second layer $(Y_2,X_1)$ are encoded by SW-coding in the presence of $X_2$ at the receiver, and finally in the last layer $Y_1$ is described using SW-coding assuming $(X_2,Y_2,X_1)$ are available to the receiver. Analyzing this strategy, yields that $(R_1,R_2)$ are achievable if, 
\begin{align*}
R_1=R_{11}+R_{12}&\ge H(X_1|X_2Y_2)+H(Y_1|X_1X_2Y_2)= H(X_1Y_1|X_2Y_2)\\
R_2=R_{21}+R_{22}&\ge H(X_2)+H(Y_2|X_1X_2)\ge H(X_2Y_2|X_1Y_1)\\        
R_1+R_2&\ge H(X_2)+H(X_1Y_2|X_2)+H(Y_1|X_2Y_2X_1)=H(X_1X_2Y_1Y_2)
\end{align*} 
From this strategy, the corner point $A$ is achieved, but the corner point $B$ is not achieved. In addition, as it can be seen in Fig. \ref{fig:SW}, the other corner point of this scheme ($C$) is coincident with one of the corner points of the two-layer scheme. By symmetry, the corner point $B$ is achieved by a three-layer scheme in which $X_1$, $(X_2,Y_1)$ and $Y_2$ are encoded in the first, second and third layer respectively. In addition, as it can be seen in Fig. \ref{fig:SW}, the union of the regions of the three different layering schemes is a closed covering of the SW-region. Note that in all the three schemes, there is a hierarchy in the sense that the first component of each source (i.e., $X_i$) is encoded prior to the second component of it (i.e., $Y_i$). 
The result of the two-terminal SW-coding suggests that to obtain the entire SW-region of multi-components DMCS, it suffices to consider all possible layering schemes such that a given hierarchy on each source is satisfied. 
\begin{definition}
An ordered partition $\mathbf{C}$ of a set $\mv$ is a sequence $[\ml_1,\ml_2,\cdots,\ml_K]$ of subsets of $\mv$, with union $\mv$, which are non-empty, and pairwise disjoint. Denote the family of all ordered partitions of a given set $\mv$, by $\mf_{\mv}$.
\end{definition}
\par
Consider a DMCS $S_{\mv}$ with two component sources, i.e., $S_v=(X_v,Y_v)$. Now we describe ML-SW with respect to a given ordered partition $\bC=[\ml_1,\cdots,\ml_K]$. In addition, we assume that the decoder has access to  side information $Z$ which is correlated with $(X_{\mv},Y_{\mv})$ according to an arbitrary distribution $p(x_{\mv},y_{\mv},z)$. 
\begin{enumerate}
\item In the first layer, using SW-coding, $X_{\ml_1}^n$ is encoded with rates $R_1=(R_{11},R_{12},\cdots,R_{1V})$ in which for $v\notin\ml_1$, we set $R_{1v}=0$. The receiver can reliably decode $X^n_{\ml_1}$ provided that 
\be
\label{eq:sw-l1}
\forall\ms\subseteq\ml_1: R_{1\ms}\ge H(X_{\ms}|X_{\ml_1\backslash\ms}Z)
\ee
Define the function $h_{\bC,1}:2^{\mv}\rightarrow \mathbb{R}$ as
\[ 
h_{\bC,1}(\ms)=H(X_{\ms\cap\ml_1}|Z)
\]
Now using the sub-modularity of the entropy function, we have 
\begin{align}
h_{\bC,1}(\ms\cap\mt)+h_{\bC,1}(\ms\cup\mt)&=H(X_{\ms\cap\mt\cap\ml_1}|
Z)+H(X_{(\ms\cup\mt)\cap\ml_1}|Z)\nonumber\\
                               &=H(X_{(\ms\cap\ml_1)\cap(\mt\cap\ml_1)}|Z)+H(X_{(\ms\cap\ml_1)\cup(\mt\cap\ml_1)}|Z)\nonumber\\
                               &\le H(X_{\ms\cap\ml_1}|Z)+H(X_{\mt\cap\ml_1}|Z)\nonumber\\
                               &=h_{\bC,1}(\ms)+h_{\bC,1}(\mt)\label{eq:sub-h}
\end{align}                               
Hence $h_{\bC,1}$ is sub-modular. In addition, we have: $h_{\bC,1}(\ms|\ms^C)=H(X_{\mv\cap\ml_1}|Z)-H(X_{\ms^C\cap\ml_1}|Z)=H(X_{\ms}|X_{\ml_1\backslash\ms},Z)$. Note that $R_{1\ms}=R_{1\ms\cap\ml_1}$, thus \eqref{eq:sw-l1} is equivalent to  
\be
\label{eq:sw-eq1}
\forall\ms\subseteq\mv: R_{1\ms}\ge h_{\bC,1}(\ms|\ms^C)
\ee 
Now it follows from Definition \ref{def:associate} that $R_1$ is contained in the SW-region of the first layer, iff it majorizes the essential polytope of $h_{\bC,1}$, i.e., $R_1\succ \mathbf{P}_{h_{\bC,1}}$.
\item In the layer $2\le i\le K+1$, assuming that $(X^n_{\ml^i},Y^n_{\ml^{i-1}})$ has been decoded at the receiver from the previous layers (where $\ml^i=\cup_{k=1}^{i-1} \ml_k$), using SW-coding $(X_{\ml_i}^n,Y^n_{\ml_{i-1}})$ is encoded with rates $R_i=(R_{i1},R_{i2},\cdots,R_{iV})$ in which for $v\notin\ml_{i-1}\cup\ml_i$, we set $R_{iv}=0$. The receiver can reliably decode $(X_{\ml_i}^n,Y^n_{\ml_{i-1}})$ provided that,
\be
\label{eq:sw-li}
\forall\ms\subseteq\ml_{i-1}\cup\ml_i: R_{i\ms}\ge H(X_{\ms\cap\ml_i}Y_{\ms\cap\ml_{i-1}}|X_{\ml_i\backslash\ms}Y_{\ml_{i-1}\backslash\ms}X_{\ml^i}Y_{\ml^{i-1}}Z)
\ee  
Define the function $h_{\bC,i}:2^{\mv}\rightarrow \mathbb{R}$ as follows:
   \[ 
      h_{\bC,i}(\ms)=H(X_{\ms\cap\ml_i}Y_{\ms\cap\ml_{i-1}}|X_{\ml^i}Y_{\ml^{i-1}}Z)
   \]
Now in similar manner to \eqref{eq:sub-h}, it can be shown that $h_{\bC,i}$ is sub-modular. Following similar steps described in the previous stage, we conclude that $R_i$ is contained in the SW-region of the layer $i$, iff it majorizes the essential polytope of $h_{\bC,i}$, i.e., $R_i\succ \mathbf{P}_{h_{\bC,i}}$.
\end{enumerate}
Define $R\triangleq\sum_{k=1}^{K+1}R_k$ (which is the overall rate vector) and $h_{\bC}\triangleq \sum_{k=1}^{K+1}h_{\bC,k}$. We showed that $R\succ \mathbf{P}_{h_{\bC}}$. On the other side, suppose that the point $R$ majorizes $\mathbf{P}_{h_{\bC}}$, so there is a point $R^*\in\mathbf{P}_{h_{\bC}}$ such that $R\succ R^*$. Applying Lemma \ref{le:sum-polytope} to $(h_{\bC,k}:1\le k\le K+1)$, we have $\mathbf{P}_{h_{\bC}}=\sum_{k=1}^{K+1}\mathbf{P}_{h_{\bC,k}}$. Hence there are points $(R^*_k\in\mathbf{P}_{h_{\bC,k}}:1\le k\le K+1)$ such that $R^*\triangleq\sum_{k=1}^{K+1}R^*_k$. Let $R_k=R^*_k+\frac{\Delta R}{K+1}$ where $\Delta R=R-R^*$. Now we have $R\triangleq\sum_{k=1}^{K+1}R_k$ and for all $k$, $R_k\succ \mathbf{P}_{h_{\bC,k}}$. Thus, each rate vector $R$ satisfying $R\succ\mathbf{P}_{h_{\bC}}$ can be achieved using ML-SW coding with respect to $\bC$. Therefore the set of all achievable rates with respect to $\bC$ is given by:
   \begin{align}
      \label{eq:sw-bc}
       \mathcal{R}_{\bC}&= \{R\in\mathbb{R}^{|\mv|}:R\succ\mathbf{P}_{h_{\bC}}\}\nonumber\\
                                    &=\{R\in\mathbb{R}^{|\mv|}:\forall \ms\subseteq\mv, R_{\ms}\ge\sum_{i=1}^{K+1} H(X_{\ms\cap\ml_i}Y_{\ms\cap\ml_{i-1}}|X_{\ml_i\backslash\ms}Y_{\ml_{i-1}\backslash\ms}X_{\ml^i}Y_{\ml^{i-1}}Z)
\}
   \end{align} 
\par The next theorem, is the main result of this section.
\begin{theorem}[SW-identity]
\label{thm:sw-covering}
The set $\{\mathcal{R}_{\bC}:\bC\in\mf_{\mv}\}$ is a closed covering of $\mr_{SW}$ which is the SW-region defined by:
   \be
      \label{eq:sw-region}
      \mr_{SW}= \{R\in\mathbb{R}^{|\mv|}:\forall \ms\subseteq\mv, R_{\ms}\ge H(X_{\ms} Y_{\ms}|X_{\ms^C}Y_{\ms^C}Z)\}
   \ee   
\end{theorem}
\begin{proof}
Define the function $h:2^{\mv}\rightarrow\mathbb{R}$ with $h({\ms})=H(X_{\ms}Y_{\ms}|Z)$. $h$ is a sub-modular function with the essential polytope $\bp_h$. By definition, a point $R$ belongs to SW-region iff it majorizes $\bp_h$. To prove the theorem, we must show that
  \be
  \label{eq:eq-thm}
    \mr_{SW}=\bigcup_{\bC\in\mf_{\mv}}\mr_{\bC}
  \ee
Applying Equation \eqref{eq:maj-property} to the RHS of \eqref{eq:eq-thm} yields, 
  \be
  \label{eq:un}
    \bigcup_{\bC\in\mf_{\mv}}\mr_{\bC}= \{R\in\mathbb{R}^{|\mv|}:R\succ\bigcup_{\bC\in\mf_{\mv}}\mathbf{P}_{h_{\bC}}\} 
  \ee 
\par Thus, to prove the theorem, we only need to show that $\{\bp_{h_{\bC}}:\bC\in\mf_{\mv}\}$ is a closed covering of $\bp_h$.
We prove this by strong induction on $|\mv|$. For
$N = 1$ as base of induction, it is clear (The case $N=2$ was proved separately in the beginning of the section). For $|\mv|\ge 2$ assume that the theorem holds for any $\mv$ with size $|\mv|\le N-1$. We show that $\{\bp_{h_{\bC}}:\bC\in\mf_{\mv}\}$ and $\bp_h$ satisfy the conditions of Lemma \ref{le:cover}, thus $\{\bp_{h_{\bC}}:\bC\in\mf_{\mv}\}$ is a closed covering of $\bp_h$.
\begin{claim}
\label{cl:1}
For any ordered partition $\bC$ of $\mv$, we have
\[  \bp_{h_{\bC}}\subseteq\bp_h.\]
\end{claim}
\emph{Proof of Claim \ref{cl:1}}.
First note that, (See equation \eqref{eq:sw-li})
    \begin{align}
    h_{\bC}(\ms|\ms^C)&=\sum_{i=1}^{K+1} H(X_{\ms\cap\ml_i}Y_{\ms\cap\ml_{i-1}}|X_{\ml_i\backslash\ms}Y_{\ml_{i-1}\backslash\ms}X_{\ml^i}Y_{\ml^{i-1}}Z) \label{eq:app-1}\\  
    &\ge \sum_{i=1}^{K+1}H(X_{\ms\cap\ml_i}Y_{\ms\cap\ml_{i-1}}|X_{\ms^C}Y_{\ms^C}X_{\ms\cap\ml^i}Y_{\ms\cap\ml^{i-1}}Z)\label{eq:p1-1}\\
    &= H(X_{\ms}Y_{\ms}|X_{\ms^C}Y_{\ms^C}Z)\label{eq:p1-2}\\
    &= h(\ms|\ms^C)\label{eq:p1-3}
   \end{align} 
where \eqref{eq:p1-1} follows from the fact that $(\ml_i\backslash\ms)\cup\ml^i\subseteq\ms^C\cup(\ms\cap\ml^i)$ with equality holds if $\ms=\mv$ and conditioning does not reduce the entropy, and \eqref{eq:p1-2} follows by the chain rule, since $\{\ml_i\cap\ms\}_{i=1}^K$ is a partition of $\ms$. Now we can conclude the claim from \eqref{eq:p1-3}. $\square$
\begin{claim}
\label{cl:2}
Suppose $\mf_{\mt^C,\mt}$ is a subset of $\mv$ that consists of all ordered partitions which are generated by concatenating an ordered partition of $\mt^C$ and an ordered partition of $\mt$, i.e., 
\[\mf_{\mt^C,\mt}=\{\bC\in\mf_{\mv}:\bC=[\bC_1,\bC_2],\bC_1\in\mf_{\mt^C}\ \mbox{and}\ \bC_2\in\mf_{\mt}\}\]
Then, the set of facets $\{\F_{h_{\bC},\mt}:\bC\in\mf_{\mt^C,\mt}\}$ is a closed covering of $\F_{h,\mt}$.
\end{claim}
\emph{Proof of Claim \ref{cl:2}}.By Lemma \ref{le:facet}, $\F_{h,\mt}$ is given by:
    \be 
        \F_{h,\mt}=\{\ux\in\mathbb{R}^{\mv}:\ux_{\mt}\in\bp_{h_1},\ux_{\mt^C}\in\bp_{h_2}\}
    \ee
In which $\bp_{h_1}$ and $\bp_{h_2}$ are the associated essential polytopes of sub-modular functions $h_1(\ms)=H(X_{\ms}Y_{\ms}|Z)$ and $h_2(\ms)=H(X_{\ms}Y_{\ms}|X_{\mt^C}Y_{\mt^C}Z)$ with domains $2^{\mt^C}$ and $2^{\mt}$, respectively. More precisely, $\bp_{h_1}$ and $\bp_{h_2}$ are given by:  
 \begin{small}
 \begin{gather*}
 \bp_{h_1}=\{\ux\in\mathbb{R}^{\mt^C}:x_{\mt^C}=H(X_{\mt^C}Y_{\mt^C}|Z)
   ,\ \mbox{and}\ \forall\ms\subset\mt^C,x_{\ms}\ge H(X_{\ms}Y_{\ms}|X_{\ms^C\cap\mt^C}Y_{\ms^C\cap\mt^C}Z)\} \\
   \bp_{h_2}=\{\ux\in\mathbb{R}^{\mt}:x_{\mt}=H(X_{\mt}Y_{\mt}|X_{\mt^C}Y_{\mt^C}Z)
   ,\ \mbox{and}\ \forall\ms\subset\mt,x_{\ms}\ge H(X_{\ms}Y_{\ms}|X_{\ms^C\cap\mt}Y_{\ms^C\cap\mt}X_{\mt^C}Y_{\mt^C}Z)\}
    \end{gather*}
 \end{small}
Now, since the size of $\mt^C$ and $\mt$ are smaller than $N$, by applying the induction assumption to essential polytopes $\bp_{h_1}$ and $\bp_{h_2}$ (with side information $\tilde{Z}=(X_{\mt^C},Y_{\mt^C},Z)$ at the decoder), we obtain:
  \begin{align}
    \bp_{h_1}&=\bigcup_{\bC_1\in\mf_{\mt^C}}\bp_{h_{1,\bC_1}}\nonumber\\
    \bp_{h_2}&=\bigcup_{\bC_2\in\mf_{\mt}}\bp_{h_{2,\bC_2}}\label{eq:p-1-1} 
  \end{align}
\par where $\bC_1=[\ml_{1,1},\cdots,\ml_{1,K_1}]$, $\bC_2=[\ml_{2,1},\cdots,\ml_{2,K_2}]$ and the functions $h_{1,\bC_1}$ and $h_{2,\bC_2}$ whose domain are $2^{\mt^C}$ and $2^{\mt}$, are defined by: 
  \begin{align}
  h_{1,\bC_1}(\ms)&=\sum_{k=1}^{K_1+1}H(X_{\ms\cap\ml_{1,k}}Y_{\ms\cap\ml_{1,k-1}}|X_{\ml_1^k}Y_{\ml_1^{k-1}}Z)\label{eq:p2-1}\\      h_{2,\bC_2}(\ms)&=\sum_{k=1}^{K_2+1}H(X_{\ms\cap\ml_{2,k}}Y_{\ms\cap\ml_{2,k-1}}|X_{\ml_2^k}Y_{\ml_2^{k-1}}\tilde{Z}) \label{eq:p2-2}
  \end{align}
Using \eqref{eq:p2-1} and \eqref{eq:p2-2}, we obtain $\bp_{h_{1,\bC_1}}$ and $\bp_{h_{2,\bC_2}}$ as:
\begin{align}
\bp_{h_{1,\bC_1}}=&\big\{\ux\in\mathbb{R}^{\mt^C}:x_{\mt^C}=H(X_{\mt^C}Y_{\mt^C}Z),\ \mbox{and}\ \forall\ms\subset\mt^C\nonumber\\ &
x_{\ms}\ge\sum_{k=1}^{K_1+1}H(X_{\ms\cap\ml_{1,k}}Y_{\ms\cap\ml_{1,k-1}}|X_{\ml_{1,k}\backslash\ms}Y_{\ml_{1,k-1}\backslash\ms}X_{\ml_1^k}Y_{\ml_1^{k-1}}Z)\big\}\label{eq:p0-1} \\ 
\bp_{h_{2,\bC_2}}=& \big\{\ux\in\mathbb{R}^{\mt}:x_{\mt}=H(X_{\mt}Y_{\mt}|\tilde{Z}),\ \mbox{and}\ \forall\ms\subset\mt\nonumber\\ &
x_{\ms}\ge\sum_{k=1}^{K_2+1}H(X_{\ms\cap\ml_{2,k}}Y_{\ms\cap\ml_{2,k-1}}|X_{\ml_{2,k}\backslash\ms}Y_{\ml_{2,k-1}\backslash\ms}X_{\ml_2^k}Y_{\ml_2^{k-1}}X_{\mt^C}Y_{\mt^C}Z)\big\}\label{eq:p0-2}
\end{align}  
Let $\bC=[\ml_{1,1},\cdots,\ml_{1,K_1},\ml_{2,1},\cdots,\ml_{2,K_2}]$ be the concatenation of $\bC_1$ and $\bC_2$. We assert that 
\be
\label{eq:p3}
\F_{h_{\bC},\mt}=\{\ux\in\mathbb{R}^{\mv}:\ux_{\mt^C}\in\bp_{h_{1,\bC_1}},\ux_{\mt}\in\bp_{h_{2,\bC_2}}\}
\ee
By Lemma \ref{le:facet}, $\ux$ belongs to $\F_{h_{\bC},\mt}$, iff
\begin{equation}
\label{eq:p4}
\begin{array}{lccr}
\ms\subseteq\mt^C:& x_{\ms}\ge h_{\bC}(\ms|\mt^C\cap\ms^C) &\mbox{with equality for}& \ms=\mt^C\\ 
\ms\subseteq\mt:& x_{\ms}\ge h_{\bC}(\ms|\ms^C) &\mbox{with equality for}& \ms=\mt
\end{array}
\ee 
To evaluate \eqref{eq:p4}, consider
\begin{align}
h_{\bC}(\ms)&=\sum_{k=1}^{K_1}H(X_{\ms\cap\ml_{1,k}}Y_{\ms\cap\ml_{1,k-1}}|X_{\ml_1^k}Y_{\ml_1^{k-1}}Z)\n&\qquad+H(X_{\ms\cap\ml_{2,1}}Y_{\ms\cap\ml_{1,K_1}}|X_{\mt^C}Y_{\ml_1^{K_1}}Z)+\sum_{k=2}^{K_2+1}H(X_{\ms\cap\ml_{2,k}}Y_{\ms\cap\ml_{2,k-1}}|X_{\ml_2^k}Y_{\ml_2^{k-1}}\tilde{Z})
\end{align}
where we have used the fact that $\ml_1^{K_1+1}=\mt^C$.
 Now, we compute the RHS of \eqref{eq:p4}:
\begin{align}
h_{\bC}(\ms|\mt^C\cap\ms^C)&=h_{\bC}(\mt^C)-h_{\bC}(\mt^C\cap\ms^C)\nonumber\\
&=\sum_{k=1}^{K_1+1}H(X_{\ml_{1,k}}Y_{\ml_{1,k-1}}|X_{\ml_1^k}Y_{\ml_1^{k-1}}Z)-H(X_{\ms^C\cap\ml_{1,k}}Y_{\ms^C\cap\ml_{1,k-1}}|X_{\ml_1^k}Y_{\ml_1^{k-1}}Z)\label{eq:p5-1}\\
&=\sum_{k=1}^{K_1+1}H(X_{\ms\cap\ml_{1,k}}Y_{\ms\cap\ml_{1,k-1}}|X_{\ml_{1,k}\backslash\ms}Y_{\ml_{1,k-1}\backslash\ms}X_{\ml_1^k}Y_{\ml_1^{k-1}}Z)\label{eq:p5-2}\\
h_{\bC}(\ms|\ms^C)&=\sum_{k=1}^{K_2+1}H(X_{\ms\cap\ml_{2,k}}Y_{\ms\cap\ml_{2,k-1}}|X_{\ml_{2,k}\backslash\ms}Y_{\ml_{2,k-1}\backslash\ms}X_{\ml_2^k}Y_{\ml_2^{k-1}}\tilde{Z})\label{eq:p5-3}
\end{align}
where \eqref{eq:p5-1} follows, because $\ms\subseteq\mt^C$ and $\mt^C$ are disjoint from all $\ml_{2,i}$, \eqref{eq:p5-3} follows from the fact that $\mt$ is disjoint from all $\ml_{1,i}$.\par
Now \eqref{eq:p0-1}, \eqref{eq:p0-2}, \eqref{eq:p5-2} and \eqref{eq:p5-3} together show the truth of assertion. Finally, the assertion with \eqref{eq:p-1-1} implies that for each point $\ux\in\F_{h,\mt}$, there exists an ordered partition $\bC\in\mf_{\mt^C,\mt}$ for which $\F_{h_{\bC},\mt}$ contains $\ux$. This completes the proof of Claim 2. $\square$
\begin{claim}
\label{cl:3}
For each facet $\F_{h_{\bC},\mt}$ of an essential polytope of a given ordered partition $\bC$ inside the $\bp_h$, there exists an ordered partition $\bC^*\neq\bC$, such that 
    \be
       \label{eq:common}
         \bp_{h_{\bC}}\bigcap\bp_{h_{\bC^*}}=\F_{h_{\bC},\mt}
    \ee  
\end{claim}
\par \emph{Proof of Claim \ref{cl:3}}. Let $\bC=[\ml_1,\cdots,\ml_K]$. From the proof of Claim \ref{cl:2}, the corresponding facets to $(\ml_{i}^{K}=\cup_{k=i}^{K}\ml_k:k\ge2)$ lie on the boundary of $\bp_h$. Thus, we only consider the facets corresponding to $\mt\neq\ml_{i}^{K}=\cup_{k=i}^{K}\ml_k$. For such $\mt$, set $\bC^*=[\ml_1^*,\cdots,\ml_K^*,\ml_{K+1}^*]$, where $\ml_k^*=(\mt\cap\ml_{k-1})\cup(\mt^C\cap\ml_k)$. Now we show that
   \be
     \label{eq:cl3}
      \F_{h_{\bC},\mt}=\F_{h_{\bC^*},\mt^C}.
   \ee    
This proves Claim \ref{cl:3}, because $\bp_{h_{\bC}}$ gets the minimum of $x_{\mt}$ on the $\F_{h_{\bC},\mt}$, and $\bp_{h_{\bC}}$ gets the maximum of $x_{\mt}$ on the $\F_{h_{\bC^*},\mt^C}$ (since $x_{\mt}=H(X_{\mv}Y_{\mv}|Z)-x_{\mt^C}$). \par
We provide the formal proof of \eqref{eq:cl3} in Appendix \ref{app:3}. Instead, we give the main idea behind the construction of $\bC^*$. First, consider the simple SW-coding of a DMCS $X_{\mv}$ with rate-tuple $R_{\mv}$. It is well-known that the minimum of $R_{\mt}$ is achieved with joint decoding of $X_{\mt^C}$ with sum-rate $R_{\mt^C}=H(X_{\mt^C}|Z)$ followed by joint decoding of $X_{\mt}$ in the presence of $X_{\mt^C}$ at the decoder with sum-rate $R_{\mt}=H(X_{\mt}|X_{\mt^C}Z)$. Also, Lemma \ref{le:facet} confirms this result about any sub-modular function. Moreover, this lemma tells us that each point which achieves the minimum of $R_{\mt}$ can be obtained by this two-level decoding. Now consider the ML-SW coding with respect to $\bC$. Each point of ML-SW region can be written in the form $R=\sum_{k=1}^{K+1}R_k$, where $R_k$ lies in the SW-region of layer $k$. So $R_{\mt}$ can be split into the rates $R_{k,\mt}=R_{k,\mt\cap(\ml_k\cup\ml_{k-1})}$. Thus to minimize $R_{\mt}$, we require to minimize each of $R_{k,\mt\cap(\ml_k\cup\ml_{k-1})}$. In layer $k$, SW-coding is done over $(X_{\ml_k},Y_{\ml_{k-1}})$, therefore to achieve the minimum of $R_{k,\mt\cap(\ml_k\cup\ml_{k-1})}$, it suffices to consider two levels of decoding at the decoder: the decoder, first decodes $(X_{\mt^C\cap\ml_k},Y_{\mt^C\cap\ml_{k-1}})$ in the presence of $(X_{\ml^k},Y_{\ml^{k-1}},Z)$, then decodes $(X_{\mt\cap\ml_k},Y_{\mt\cap\ml_{k-1}})$ in the presence of $(X_{\ml^k},Y_{\ml^{k-1}},X_{\mt^C\cap\ml_k},Y_{\mt^C\cap\ml_{k-1}},Z)$. In overall, to minimize $R_{\mt}$ with respect to $\bC$, one can consider the following $2K+2$ levels of decoding:
  \be
  \label{eq:chain-1}   X_{\mt^C\cap\ml_1},X_{\mt\cap\ml_1},\cdots,(X_{\mt^C\cap\ml_k},Y_{\mt^C\cap\ml_{k-1}}),(X_{\mt\cap\ml_k},Y_{\mt\cap\ml_{k-1}}),\cdots,Y_{\mt^C\cap\ml_{K}},Y_{\mt\cap\ml_K}
  \ee 
On the other side, to maximize $R_{\mt}$ (or equivalently, to minimize $R_{\mt^C}$) with respect to $\bC^{*}$, the following order on SW-coding is required,
 \be
  \label{eq:chain-2}   X_{\mt\cap\ml_1^*},X_{\mt^C\cap\ml_1^*},\cdots,(X_{\mt\cap\ml_k^*},Y_{\mt\cap\ml_{k-1}^*}),(X_{\mt^C\cap\ml_k^*},Y_{\mt^C\cap\ml_{k-1}^*}),\cdots,Y_{\mt\cap\ml_{K+1}^*},Y_{\mt^C\cap\ml_{K+1}^*}
  \ee
  Now, note that $\mt^C\cap\ml_k^*=\mt^C\cap((\mt\cap\ml_{k-1})\cup(\mt^C\cap\ml_k))=\mt^C\cap\ml_k$ and $\mt\cap\ml_k^*=\mt\cap((\mt\cap\ml_{k-1})\cup(\mt^C\cap\ml_k))=\mt\cap\ml_{k-1}$; in particular $\mt\cap\ml_1^*=\mt^C\cap\ml_{K+1}^*=\emptyset$. Comparing \eqref{eq:chain-1} with \eqref{eq:chain-2}, we see that these two multi-level decoding schemes are the same, thus the intersection of $\bp_{h_{\bC}}$ and $\bp_{h_{\bC*}}$ is $\F_{h_{\bC},\mt}=\F_{h_{\bC^*,\mt^C}}$. $\square$
\par Now Claims \ref{cl:1}--\ref{cl:3} ensure that $\bp_{h_{\bC}}$ and $\bp_h$ satisfy the conditions of Lemma \ref{le:cover}. This completes the proof.                                 
\end{proof}
\section{Feasible Constraints for reliable multicasting of a DMCS over cooperative networks}\label{sec:5}
In this section, we obtain a set of DMCS which can reliably be multicast over a cooperative network. Our approach is based on generalization of the CF strategy for relay networks. Two types of generalization have been considered in the previous works, \cite{kramer2005,rost}. In \cite{kramer2005}, the CF strategy was generalized in the following manner: 
\begin{enumerate}
\item Each relay and destination partially decode the messages of the other relays.
\item Each relay compresses its observation $Y_v$, in the presence of side information from messages of the other relays.
\item Each relay sends its compressed observation through a Multiple Access Channel (MAC). Finally, destination decodes the source message.
\end{enumerate}
This scenario deals with relays in a symmetric way, i.e., all relays lie in a single MAC layer. In \cite{rost}, a generalization of \emph{mixed strategy} of \cite[Theorem 7]{cover} is proposed. By relaxing the partial decode-and-forward part of the mixed strategy, we obtain a generalization of the CF strategy. In this scenario, relays are ordered according to a given permutation. Each relay compresses its observation using multiple description method (MD) and sends these descriptions through a broadcast channel with a degraded message set. Each relay and destination decode their respective descriptions after decoding their broadcast messages according to a sequential decoding scheme. However, if the relays use the simple Wyner-Ziv coding rather than MD, the result is a special case of \cite[Theorem 3]{kramer2005}. In another scenario proposed in \cite{rost:asilomar}, CF is generalized for half-duplex channels. Although, this method is proposed for half-duplex relay networks, it can be generalized for relay networks, too. In this scenario, each relay uses the simple Wyner-Ziv coding. This scenario differs from the previous generalization of CF, in which the destination considers an ordering of relays, and decodes the compressed observation of relay $k$, in the presence of compressed observations of relays $(k+1,k+2,\cdots,N-1)$ which are decoded in the previous blocks. This is similar to ML-SW coding.  \par
 We propose a joint source coding and Wyner-Ziv coding for multicasting a DMCS $U_{\ma}$ over cooperative networks. In this scenario, in each block, each node compresses its observation using Wyner-Ziv coding, then in the next block jointly maps the compressed observation and its current source sequence to a channel input codeword and transmits the codeword. The joint encoding used in this scheme, benefits from the advantage of joint source-channel coding in comparison with source-channel separation in the multicast scenario, which is illustrated in \cite{tuncel}. Moreover, in this scheme, each node has two types of sources including the compressed observation and the source sequence which are required to decode at each destination. By the nature of relaying, it is not possible to decode these sources, simultaneously. This situation is similar to ML-SW coding, in which two components of the source are not being decoded simultaneously. Motivated by the results of ML-SW coding, e.g., Theorem \ref{thm:sw-covering}, each destination groups the other nodes into some layers according to its ability to decode the information of other nodes. Using insights from the ML-SW, in the first level of decoding, the destination can directly decode the first component of the information of nodes in the first layer, i.e., the source sequences of the first layer, through a MAC layer between layer one and the destination, and in  level $k$ of decoding, destination decodes the source sequences of layer $k$ and the compressed observations of layer $k-1$ (second component of information of  layer $k-1$) jointly through the MAC layer between layer $k$ and the destination in the presence of the decoded information from levels $(1,2,\cdots,k-1)$ as side information. These side information play two roles in improving  the decoding:
 \begin{enumerate}
 \item These are side information for Slepian-Wolf coding that enlarge the SW-region.
  \item  These are side information for MAC that enlarge the MAC-region. Unlike the first role, this role does not arise from the ML-SW.
  \end{enumerate}
\par  Enlarging the SW-region and the MAC-region provides the opportunity for some intersection between the two regions which results in the reliable transmission of source sequences of the nodes in layer $k$ and the compressed observations of the nodes in layer $k-1$ in an operational separation sense, even if the original MAC region does not intersect with the original SW-region. \par
  The next theorem is the main result of the paper.  
    \begin{theorem}
\label{thm:sw}
The set of DMCS $U_{\ma}$ can reliably be multicast over a cooperative network to nodes in $\md$, if there exist auxiliary random variables $\hY_{\mv}$ and $Q$, such that for each $\ms\subseteq\ma$, we have 
   \be
     \label{eq:sw}
      H(U_{\ms}|U_{\ma\backslash\ms})<  \min_{d_i\in\md\backslash\ms}\min_{\mv\supseteq\mw\supseteq\ms: \atop d_i\in\mw^C} [I(X_{\mw};Y_{d_i}\hY_{\mw^C\backslash\{d_i\}}|X_{\mw^C}Q)-I(Y_{\mw};\hY_{\mw}|X_{\mv}Y_{d_i}\hY_{\mw^C\backslash\{d_i\}}Q)]
 \ee
  \noindent where the joint p.m.f. of random variables 
  factors as
 \be
 \label{eq:dist}
 p(q)p(u_{\ma})[\prod_{v\in\mv}p(x_v|q)p(\hy_v|x_v,y_v,q)]p(y_{\mv}|x_{\mv}).
 \ee
 \end{theorem}
 \begin{remark}
 The constraint \eqref{eq:sw} separates source coding from channel coding in the operational separation sense \cite{tuncel}.
 To see this, observe that the constraint \eqref{eq:sw} is equivalent to the following constraint,
 \be\label{eq:sw-5}
 \forall\mw\subseteq\mv,d_i\in\mw^C: H(U_{\mw\cap\ma}|U_{\ma\backslash\mw})+I(Y_{\mw};\hY_{\mw}|X_{\mv}Y_{d_i}\hY_{\mw^C\backslash\{d_i\}}Q)<I(X_{\mw};Y_{d_i}\hY_{\mw^C\backslash\{d_i\}}|X_{\mw^C}Q).
 \ee
 Consider a cut $\Lambda=(\mw,\mw^C)$. The RHS of the \eqref{eq:sw-5} provides an achievable flow through the cut $\Lambda$. The first term in the LHS of \eqref{eq:sw-5} represents the rate of the Slepian-Wolf coding for describing $U_{\mw\cap\ma}$ to the destinations in the other side of the cut in the presence of $U_{\ma\backslash\mw}$ which is available in $\mw^C$. The second term in the LHS of \eqref{eq:sw-5} can be interpreted as the rate of the Wyner-Ziv coding for describing a compression of the observation $Y_{\mw}$, i.e.  $\hY_{\mw}$, to the other side of the cut in the presence of $(X_{\mw^C},\hY_{\mw^C},Y_{d_i})$ and $X_{\mw}$, which the latter can be regarded as the output of channel decoder. Since the compression rate of the sources is less than the information flow, one can expect that the multicasting of the sources is feasible, due to the source-channel separation approach.  \end{remark} 
 \begin{proof}[Proof of Theorem \ref{thm:sw}] 
For the sake of simplicity, we assume that $|\mq|=1$ where $Q$ is a time-sharing random variable. First, we characterize a set of DMCS which can reliably be multicast over a cooperative network, with respect to  given ordered partitions at each destination. For each destination node $d_i$, let $\mv_{-d_i}=\mv\backslash\{d_i\}$. The following lemma, establishes a set of sufficient conditions for reliable multicasting of $U_{\ma}$ over the cooperative network. We provide the proof of it in Subsection \ref{sub:a}.   
\begin{lemma}
\label{le:first}
The set of DMCS $U_{\ma}$ can reliably be multicast over a cooperative network to subset $\md$ of the nodes, if for each $d_i\in\md$, there exists an ordered partition $\mathbf{C}^{(d_i)}=[\ml_1,\ml_2,\cdots,\ml_{\ell}]$ of $\mv_{-d_i}$ such that for each $\ms\subseteq\mv_{-d_i}$, the following constraint is satisfied:   
\begin{align}
\label{eq:sufficient}
\sum_{t\in\ms}H(X_t)+H(\hY_t|X_tY_t)\geq&
\sum_{k=1}^{\ell+1}\Big(H(U_{\ms\cap\ml_k}|U_{\ml_k\backslash\ms}U_{\ml^k}U_{d_i})+\nonumber\\
&\quad\qquad H(X_{\ms\cap\ml_k}\hY_{\ms\cap\ml_{k-1}}|X_{\ml_k\backslash\ms}\hY_{\ml_{k-1}\backslash\ms}X_{\ml^k}\hY_{\ml^{k-1}}Y_{d_i}X_{d_i})\Big),
 \end{align}
 \noindent where the random variables $(x_{\mv},y_{\mv},\hy_{\mv})$ are distributed according to \eqref{eq:dist}. 
\end{lemma}
\par This lemma gives a partial solution to the problem of reliable multicasting of $U_{\ma}$ over the cooperative network, in the sense that to find out that the multicasting of $U_{\ma}$ is feasible, we must consider all the possible ordered partitions of each set $\mv_{-d_i}$ and check that the constraint \eqref{eq:sufficient} is satisfied or not. If for each destination node $d_i$, there exists at least one ordered partition of $\mv_{-d_i}$ such that \eqref{eq:sufficient} is satisfied, then reliable multicasting is feasible. Since the number of ordered partitions of a set $\mv$ grows rapidly with $|\mv|$, such approach (checking the constraint \eqref{eq:sufficient} for all the ordered partitions) seems to be difficult. However, using Theorem \ref{thm:sw-covering}, we show that there exists a set of constraints that unifies the set of constraints \eqref{eq:sufficient} with respect to all the ordered partitions. The following lemma, establishes such a result.
\begin{lemma}[Unified Sufficient Conditions]\label{le:6}
For a destination node $d_i$, there exists at least one ordered partition $\bC^{(d_i)}$ of $\mv_{-d_i}$ for which the constraint \eqref{eq:sufficient} is satisfied, if and only if the following constraint is satisfied, 
\be
\label{eq:uniuni}
\forall \ms\subseteq\mv_{-d_i}:\sum_{t\in\ms}H(X_t)+H(\hY_t|X_tY_t)\geq
 H(\hY_{\ms}X_{\ms}|X_{\ms^C}\hY_{\ms^C}Y_{d_i}X_{d_i})+H(U_{\ms}|U_{\ms^C}U_{d_i}).
\ee
\end{lemma}
\emph{Proof of Lemma \ref{le:6}}.  For each $v\in\mv$, define $R_v=H(X_v)+H(\hY_v|Y_vX_v)$ and
\[R^{(d_i)}=(R_1,\cdots,R_{d_i-1},R_{d_i+1},\cdots,R_V).\]
 Consider the RHS of \eqref{eq:sufficient}. Since random variables $U_{\ma}$ and $(X_{\mv},\hY_{\mv},Y_{\mv})$ are independent, the constraint \eqref{eq:sufficient} can be rewritten as
 \be\label{eq:correc}
 \forall\ms\subseteq\mv_{-d_i}: R^{(d_i)}_{\ms}\ge \sum_{k=1}^{\ell+1}H(U_{\ms\cap\ml_k}X_{\ms\cap\ml_k}\hY_{\ms\cap\ml_{k-1}}|U_{\ml_k\backslash\ms}X_{\ml_k\backslash\ms}\hY_{\ml_{k-1}\backslash\ms}U_{\ml^k}X_{\ml^k}\hY_{\ml^{k-1}}U_{d_i}Y_{d_i}X_{d_i}).
 \ee
 The RHS of \eqref{eq:correc} can be expressed in the form of \eqref{eq:sw-bc} with $\mv=\mv_{-d_i}$ , $X_v=(X_v,U_v)$, $Y_v=\hY_v$ and $Z=(Y_{d_i},X_{d_i},U_{d_i})$, thus the constraint \eqref{eq:sufficient} is equivalent to  $R^{(d_i)}\in\mr_{\bC^{(d_i)}}$. Therefore for the node $d_i$, there exists at least one ordered partition of $\mv_{-d_i}$ such that \eqref{eq:sufficient} is satisfied, iff $R^{(d_i)}\in\cup_{\bC^{(d_i)}\in\mf_{\mv_{-d_i}}}\mr_{\bC^{(d_i)}}$. Applying Theorem \ref{thm:sw-covering}, we conclude that such $\mathbf{C}^{(d_i)}$ exists iff \eqref{eq:uniuni} is satisfied.$\square$
\par The constraint \eqref{eq:uniuni} can be rewritten in the following form:
\begin{subequations} \label{eq:adi}\begin{align}
\forall  \ms\subseteq\ma\backslash\{d_i\} :&
H(U_{\ms}|U_{\ma\backslash\ms})\le \min_{\mw\supseteq\ms\atop d_i\in\mw^C} R^{(d_i)}_{\mw}-H(\hY_{\mw}X_{\mw}|X_{\mw^C}\hY_{\mw^C\backslash\{d_i\}}Y_{d_i}),\label{eq:adi1}\\ \label{eq:adi2}
\forall \ms\subseteq\ma^C\backslash\{d_i\}:
&R^{(d_i)}_{\ms} - H(\hY_{\ms}X_{\ms}|X_{\ms^C}\hY_{\ms^C\backslash\{d_i\}}Y_{d_i})\ge 0 .
\end{align}\end{subequations}
Consider the constraint \eqref{eq:adi}. In Appendix \ref{app:simplify}, using the joint p.m.f. \eqref{eq:dist} we will show that this constraint is equivalent to the following constraint
\begin{subequations} \label{eq:adi-adi}\begin{align}
\forall  \ms\subseteq\ma\backslash\{d_i\} :&
 H(U_{\ms}|U_{\ma\backslash\ms})<  \min_{\mw\supseteq\ms: \atop d_i\in\mw^C} [I(X_{\mw};Y_{d_i}\hY_{\mw^C\backslash\{d_i\}}|X_{\mw^C})-I(Y_{\mw};\hY_{\mw}|X_{\mv}Y_{d_i}\hY_{\mw^C\backslash\{d_i\}})]
,\label{eq:adi-adi1}\\ \label{eq:adi-adi2}
\forall \ms\subseteq\ma^C\backslash\{d_i\}:
& I(\hY_{\ms};Y_{\ms}|X_{\mv}\hY_{\ms^C\backslash\{d_i\}}Y_{d_i})\le I(X_{\ms};\hY_{\ms^C\backslash\{d_i\}}Y_{d_i}|X_{\ms^C}).
\end{align}\end{subequations}
The first constraint \eqref{eq:adi-adi1} is the same as the constraint \eqref{eq:sw}, so we only need to show that the second constraint \eqref{eq:adi-adi2} is an additional constraint. The second constraint represents a sufficient condition for reliable multicasting of the compressed observations of the non-source nodes to the destinations. Since the destinations only need to decode the sources and do not need to decode any other information, it is logical to neglect the second constraint, which completes the proof of Theorem \ref{thm:sw}. We provide a rigorous proof of this fact in subsection \ref{sub:c}.  
\end{proof}

 \subsection{Multi-Layer Slepian-Wolf coding over a cooperative network (Proof of Lemma \ref{le:first})}
\label{sub:a}
 We transmit $s_{nB}=nB$-length source over cooperative network in $B+2V-3$ blocks of length $n$ where $V$ is the cardinality of $\mv$. Observe that $r_{nB}=n(B+2V-3)$ and $\dfrac{r_{nB}}{s_{nB}}\rightarrow 1$ as $B\rightarrow \infty$, thus the sequence $\{(s_{nB},r_{nB})\}_{B=1}^{\infty}$ satisfies the condition of Definition \ref{def:sw}.
\subsubsection*{Codebook generation at node $v$}  
Fix $0<\epsilon''<\epsilon'<\epsilon$. Also fix $\delta>0$ such that $|\styp(U_v)|<2^{n(H(U_{v})+\delta)}$. To each element of $\styp(U_v)$, assign a number $w_v\in[1,2^{n(H(U_{v})+\delta)}]$ using a one-to-one mapping. Moreover, for each non-typical sequence, set $w_v =1$. Denote the result by $\uu_{v}(w_v)$. For channel coding,  independently repeat the following procedure $V$ times. Denote the resulting $k$-th codebook by $\mc_v(k)$.\par 
 Choose $2^{n(H(U_{v})+I(Y_v;\hY_v|X_v)+2\delta)}$ codewords $\ux_v(w_v,z_v)$, each drawn uniformly and independently from the set $\stypt(X_v)$ where $z_v\in[1,2^{n(I(Y_v;\hY_v|X_v)+\delta)}]$. For Wyner-Ziv coding, for each $\ux_v(w_v,z_v)$ choose $2^{n(I(Y_v;\hY_v|X_v)+\delta)}$ codewords $\uhy_v(z'_v|\ux_v)$, each drawn uniformly and independently from the set $\stypo(\hY_v|\ux_v)$ where $z'_v\in[1,2^{n(I(Y_v;\hY_v|X_v)+\delta)}]$. 
\subsubsection*{Encoding at node $v$}
Divide the $nB$-length source stream $u_v^{nB}$ into $B$ vectors $(\uu_{v,[j]}:1\leq j\leq B)$ where $\uu_{v,[j]}=(u_{v,(j-1)n+1},\cdots,u_{v,jn})$. We say that the channel encoder receives $\um_v=(m_{v,[1]},\cdots,m_{v,[B]})$, if for $1\leq j\leq B$, $\uu_{v,[j]}$ is assigned to $m_{v,[j]}\in[1,2^{n(H(U_{v})+\delta)}]$. Encoding is performed in $B+2V-3$ blocks where in block $b$, we use the codebook $\mc_v(b\mod V)$. For $1\leq b\leq B+2V-3$, define:
\[
 w_{v,[b]}= \left\{
            \begin{array}{ll}
                m_{v,[b-V+1]} &, V\le b\le B+V-1\\
                1 & ,\mbox{otherwise}.
            \end{array}\right.
            \]
 \par In block $1$, a default codeword, $\ux_v(1,1)$ is transmitted. In block $b>1$, knowing $z_{v,[b-1]}$ from Wyner-Ziv coding at the end of block $b-1$ (described below), node $v$ transmits $\ux_v(w_{v,[b]},z_{v,[b-1]})$.
 
 \subsubsection*{Wyner-Ziv coding} 
 At the end of block $b$, node $v$ knows $(\ux_{v,[b-1]},\uy_{v,[b-1]})$ and declares that $z_{v,[b-1]}=z_v$ is received if $z_{v}$ is the smallest index such that $(\uhy_{v,[b-1]}(z_{v}|\ux_{v,[b-1]}),\ux_{v,[b-1]},\uy_{v,[b-1]})$ are jointly typical. Since we have more than $2^{nI(Y_v;\hY_v|X_v)}$ codewords, such a $z_v$ exists with high probability. (See Table \ref{ta:enc} which illustrates encoding for a network with four nodes in which node $4$ is only a destination, i.e., $\mu_4=\mx_4=\emptyset$.)
\begin{table*}

\centering
    \caption{Encoding Scheme for Multicasting of two blocks of source sequences over a network with $\mv=\{1,2,3,4\}$, $\ma=\{1,2,3\}$, $\md=\{3,4\}$ and node $4$ has no channel input, i.e., $\mu_4=\mx_4=\emptyset$.}
    \label{ta:enc}
    \vspace{-0.6cm}
     \resizebox{\textwidth}{!} {%
           \begin{tabular}[t]{|c|c|c|c|c|c|c|c|}
               \hline
               Node &Block 1& Block 2& Block 3& Block 4& Block 5& Block 6& Block 7 \\
                \hline \hline
                &  &  &  & $\uu_1(m_{1[1]})$ & $\uu_1(m_{1[2]})$ &  &  \\
                 1 & $\ux_1(1,1)$ & $\ux_1(1,z_{1[1]})$ & $\ux_1(1,z_{1[2]})$ & $\ux_1(m_{1[1]},z_{1[3]})$ & $\ux_1(m_{1[2]},z_{1[4]})$ & $\ux_1(1,z_{1[5]})$ & $\ux_1(1,z_{1[6]})$ \\
                & $\uhy_1(z_{1[1]}|\ux_{1[1]})$ & $\uhy_1(z_{1[2]}|\ux_{1[2]})$ & $\uhy_1(z_{1[3]}|\ux_{1[3]})$ & $\uhy_1(z_{1[4]}|\ux_{1[4]})$ & $\uhy_1(z_{1[5]}|\ux_{1[5]})$ & $\uhy_1(z_{1[6]}|\ux_{1[6]})$ & $\uhy_1(z_{1[7]}|\ux_{1[7]})$\\
                \hline
                 &  &  &  & $\uu_2(m_{2[1]})$ & $\uu_2(m_{2[2]})$ &  &  \\
                 2 & $\ux_2(1,1)$ & $\ux_2(1,z_{2[1]})$ & $\ux_2(1,z_{2[2]})$ & $\ux_2(m_{2[1]},z_{2[3]})$ & $\ux_2(m_{2[2]},z_{2[4]})$ & $\ux_2(1,z_{2[5]})$ & $\ux_2(1,z_{2[6]})$ \\
                & $\uhy_2(z_{2[1]}|\ux_{2[1]})$ & $\uhy_2(z_{2[2]}|\ux_{2[2]})$ & $\uhy_2(z_{2[3]}|\ux_{2[3]})$ & $\uhy_2(z_{2[4]}|\ux_{2[4]})$ & $\uhy_2(z_{2[5]}|\ux_{2[5]})$ & $\uhy_2(z_{2[6]}|\ux_{2[6]})$ & $\uhy_2(z_{2[7]}|\ux_{2[7]})$\\
                \hline
                &  &  &  & $\uu_3(m_{3[1]})$ & $\uu_3(m_{3[2]})$ &  &  \\
                3 & $\ux_3(1,1)$ & $\ux_3(1,z_{3[1]})$ & $\ux_3(1,z_{3[2]})$ & $\ux_3(m_{3[1]},z_{3[3]})$ & $\ux_3(m_{3[2]},z_{3[4]})$ & $\ux_3(1,z_{3[5]})$ & $\ux_3(1,z_{3[6]})$ \\
                & $\uhy_3(z_{3[1]}|\ux_{3[1]})$ & $\uhy_3(z_{3[2]}|\ux_{3[2]})$ & $\uhy_3(z_{3[3]}|\ux_{3[3]})$ & $\uhy_3(z_{3[4]}|\ux_{3[4]})$ & $\uhy_3(z_{3[5]}|\ux_{3[5]})$ & $\uhy_3(z_{3[6]}|\ux_{3[6]})$ & $\uhy_3(z_{3[7]}|\ux_{3[7]})$ \\ 
   \hline
       \end{tabular}}
\end{table*}

\subsubsection*{Decoding at node $d_i$}
Let $\bC^{(d_i)}=[\ml_1,\cdots,\ml_{\ell}]$ be an ordered partition of the set $\mv_{-d_i}=\mv\backslash\{d_i\}$. We propose a sliding window decoding with respect to $\bC^{(d_i)}$. Define $s_{v,[b]}=(w_{v,[b]},z_{v,[b-1]})$. Suppose that $(s_{\ml_1,[b-1]},s_{\ml_2,[b-2]},\cdots,s_{\ml_\ell,[b-\ell]})$ have been correctly decoded  at the end of block $b-1$. Node $d_i$, declares that $(\hat{s}_{\ml_1,[b]},\cdots,\hat{s}_{\ml_{\ell},[b-\ell+1]})$ has been sent, if it is a unique tuple such that for each $1\le k\le\ell+1$ satisfies the following conditions,
\begin{small}
\be
\label{eq:typ}
\begin{array}{lc}
\begin{split} 
  \Big(\ux_{\ml_k}(\hat{s}_{\ml_k,[b-k+1]}),\uhy_{\ml_{k-1}}(\hat{z}_{\ml_{k-1},[b-k+1]}|\ux_{\ml_{k-1},[b-k+1]}),\ux_{\ml^k,[b-k+1]},&\\ 
  \uhy_{\ml^{k-1},[b-k+1]},\uy_{d_i,[b-k+1]},\ux_{d_i,[b-k+1]}\Big)\in\styp,&\quad\mbox{for all $k$ such that $k\le b $}
 \\
(\uu_{\ml_k}(\hat{w}_{\ml_k,[b-k+1]}),\uu_{\ml^k}(w_{\ml^k,[b-k+1]}), \uu_{d_i}(w_{d_i,[b-k+1]}))\in\styp,&\quad \mbox{for all $k$ such that $V\le b-k+1\le V+B-1$}
\end{split}
\end{array}
\ee
\end{small}
\noindent where $\hat{s}_{\ml_k,[b-k+1]}=(\hat{w}_{\ml_k,[b-k+1]},\hat{z}_{\ml_k,[b-k]})$. Note that at the end of block $b+V+\ell-2$, the vector $w_{\ma,[b+V-1]}=m_{\ma,[b]}$ is decoded. Since each $(\uu_{v,[b]}:v\in\ma)$ is jointly typical with high probability, we find the source sequence $\uu_{\ma,[b]}$ with small probability of error. Hence at the end of block $B+V+\ell-2$, $u_{\ma}^{nB}$ is decoded with small probability of error. \par

Note that in the first $V-1$ blocks of decoding, no sources information is decoded. The advantage of decoding compressed observation in these blocks is to provide side information at the receiver, in order to improve decoding in the next blocks.   
\begin{table*}

\centering
    \caption{Illustration of decoding scheme of four-node network depicted in Table \ref{ta:enc}, at node $4$ with respect to the ordered partition $\bC^{(4)}=[\{1,2\},\{3\}]$ at the end of blocks $2$ and $5$. The gray cells highlight the random variables corresponding to the unknown indices and the yellow cells highlight the random variables available at decoder  which will be used for decoding of the unknown indices through a joint typicality condition between them and the gray random variables.}
    \label{ta:dec}
    \vspace{-0.6cm}
     \resizebox{\textwidth}{!} {%
           \begin{tabular}[t]{|c|c|c||c|c|c|c|c|}
               \hline
               Node &Block 1& Block 2& Block 3& Block 4& Block 5& Block 6& Block 7 \\
                \hline \hline
                &  &  &  &\cellcolor{gray!20!yellow} $\uu_1(m_{1[1]})$ & \cellcolor{gray!30}$\uu_1(m_{1[2]})$ &  &  \\
                 1 &\cellcolor{gray!20!yellow} $\ux_1(1,1)$ & $\cellcolor{gray!30}\ux_1(1,z_{1[1]})$&\cellcolor{gray!20!yellow} $\ux_1(1,z_{1[2]})$ & \cellcolor{gray!20!yellow}$\ux_1(m_{1[1]},z_{1[3]})$ & \cellcolor{gray!30}$\ux_1(m_{1[2]},z_{1[4]})$ & $\ux_1(1,z_{1[5]})$ & $\ux_1(1,z_{1[6]})$ \\
                & \cellcolor{gray!30}$\uhy_1(z_{1[1]}|\ux_{1[1]})$ & $\uhy_1(z_{1[2]}|\ux_{1[2]})$ &\cellcolor{gray!20!yellow} $\uhy_1(z_{1[3]}|\ux_{1[3]})$ & \cellcolor{gray!30}$\uhy_1(z_{1[4]}|\ux_{1[4]})$ & $\uhy_1(z_{1[5]}|\ux_{1[5]})$ & $\uhy_1(z_{1[6]}|\ux_{1[6]})$ & $\uhy_1(z_{1[7]}|\ux_{1[7]})$\\
                \hline
                 &  &  &  & \cellcolor{gray!20!yellow}$\uu_2(m_{2[1]})$ &\cellcolor{gray!30} $\uu_2(m_{2[2]})$ &  &  \\
                 2 &\cellcolor{gray!20!yellow} $\ux_2(1,1)$ & \cellcolor{gray!30}$\ux_2(1,z_{2[1]})$ &\cellcolor{gray!20!yellow} $\ux_2(1,z_{2[2]})$ & \cellcolor{gray!20!yellow}$\ux_2(m_{2[1]},z_{2[3]})$ &\cellcolor{gray!30} $\ux_2(m_{2[2]},z_{2[4]})$ & $\ux_2(1,z_{2[5]})$ & $\ux_2(1,z_{2[6]})$ \\
                & \cellcolor{gray!30}$\uhy_2(z_{2[1]}|\ux_{2[1]})$ & $\uhy_2(z_{2[2]}|\ux_{2[2]})$ &\cellcolor{gray!20!yellow} $\uhy_2(z_{2[3]}|\ux_{2[3]})$ & \cellcolor{gray!30}$\uhy_2(z_{2[4]}|\ux_{2[4]})$ & $\uhy_2(z_{2[5]}|\ux_{2[5]})$ & $\uhy_2(z_{2[6]}|\ux_{2[6]})$ & $\uhy_2(z_{2[7]}|\ux_{2[7]})$\\
                \hline
                &  &  &  & \cellcolor{gray!30}$\uu_3(m_{3[1]})$ & $\uu_3(m_{3[2]})$ &  &  \\
                3 & $\ux_3(1,1)$ & $\ux_3(1,z_{3[1]})$ &\cellcolor{gray!20!yellow} $\ux_3(1,z_{3[2]})$ & \cellcolor{gray!30}$\ux_3(m_{3[1]},z_{3[3]})$ & $\ux_3(m_{3[2]},z_{3[4]})$ & $\ux_3(1,z_{3[5]})$ & $\ux_3(1,z_{3[6]})$ \\
                & $\uhy_3(z_{3[1]}|\ux_{3[1]})$ & $\uhy_3(z_{3[2]}|\ux_{3[2]})$ &\cellcolor{gray!30} $\uhy_3(z_{3[3]}|\ux_{3[3]})$ & $\uhy_3(z_{3[4]}|\ux_{3[4]})$ & $\uhy_3(z_{3[5]}|\ux_{3[5]})$ & $\uhy_3(z_{3[6]}|\ux_{3[6]})$ & $\uhy_3(z_{3[7]}|\ux_{3[7]})$ \\ 
   \hline & &&& \cellcolor{gray!20!yellow}$\uu_{4[4]}$ & \cellcolor{gray!20!yellow}$\uu_{4[5]}$ &  &\\
   4& $\cellcolor{gray!20!yellow}\uy_{4[1]}$ & \cellcolor{gray!20!yellow}$\uy_{4[2]}$ &\cellcolor{gray!20!yellow} $\uy_{4[3]}$ & \cellcolor{gray!20!yellow}$\uy_{4[4]}$ & \cellcolor{gray!20!yellow}$\uy_{4[5]}$ & $\uy_{4[6]}$ & $\uy_{4[7]}$\\
   \hline
   Decoding& & & & $\hat{m}_{\{1,2\},[1]}$&\cellcolor{gray!30}$\hat{m}_{\{1,2\},[2]}$,$\hat{m}_{3,[1]}$ &$\hat{m}_{3,[2]}$ &\\
   at node 4& $\emptyset$& \cellcolor{gray!30}$\hat{z}_{\{1,2\},[1]}$&$\hat{z}_{\{1,2\},[2]}$,$\hat{z}_{3,[1]}$&$\hat{z}_{\{1,2\},[3]}$,$\hat{z}_{3,[2]}$&\cellcolor{gray!30}$\hat{z}_{\{1,2\},[4]}$,$\hat{z}_{3,[3]}$&$\hat{z}_{\{1,2\},[5]}$,$\hat{z}_{3,[4]}$&$\hat{z}_{\{1,2\},[6]}$,$\hat{z}_{3,[5]}$\\
   \hline
       \end{tabular}}
\end{table*} 
\begin{example} Consider the four-node network of Table \ref{ta:enc}. Here we assume that node $4$ observes source $U_4$ correlated with the other sources. Let $\bC^{(4)}=[\{1,2\},\{3\}]$. Decoding at node $4$ begins at the end of block $2$. In block $2$, node $4$ declares that $\hat{z}_{\{1,2\},[1]}$ is decoded if $(\ux_1(1,\hat{z}_{1,[1]}),\ux_2(1,\hat{z}_{2,[1]}),\uy_{4[2]})$ and \\ $(\uhy_1(\hat{z}_{1,[1]}|\ux_{1[1]}),\uhy_2(\hat{z}_{2,[1]}|\ux_{2[1]}),\ux_{1[1]}(1,1),\ux_{2[1]}(1,1),\uy_{4[1]})$ are jointly typical. In the next block, $(\hat{z}_{\{1,2\},[2]},\hat{z}_{3,[1]})$ are decoded and in block $b$, $(\hat{w}_{\{1,2\},[b]},\hat{z}_{\{1,2\},[b-1]},\hat{w}_{3,[b-1]},\hat{z}_{3,[b-2]})$ are decoded, if (See Table \ref{ta:dec})
\begin{align}
(\uu_{\{1,2\}}(\hat{w}_{\{1,2\},[b]}),\uu_{4}(w_{4[b]}))&\in\styp\\
(\ux_{\{1,2\}}(\hat{w}_{\{1,2\},[b]},\hat{z}_{\{1,2\},[b-1]}),\uy_{4[b]})&\in\styp\\
(\uu_{3}(\hat{w}_{3,[b-1]}),\uu_{\{1,2\},[b-1]},\uu_{4}(w_{4[b-1]}))&\in\styp\\
(\ux_{3}(\hat{w}_{3,[b-1]},\hat{z}_{3,[b-2]}),\uhy_{\{1,2\}}(\hat{z}_{\{1,2\},[b-1]}|\ux_{\{1,2\}[b-1]}),\ux_{\{1,2\}[b-1]},\uy_{4[b-1]})&\in\styp\\
(\uhy_{3}(\hat{z}_{3,[b-2]}|\ux_{3[b-2]}),\ux_{\{1,2,3\}[b-2]},\uhy_{\{1,2\}[b-2]},\uy_{4[b-2]})&\in\styp
\end{align}
\end{example}

\subsubsection*{ Error Probability Analysis}
Let $\mathbf{U}_{v,[b-V+1]}$ be the observed sequence at node $v$, which is used for encoding in block $b$. We bound the error  probability of decoding at the end of block $b$ averaged over $(\bU_{\ma[b-V+1]},\bU_{\ma[b-V]},\cdots,\bU_{\ma[b-\ell-V+2]})$ and all random codebooks, assuming that no error occurred in the decoding of the previous blocks. Let $S_{v[j]}=(W_{v[j]},Z_{v[j-1]})$, in which $W_{v[j]}$ and $Z_{v[j-1]}$ are the indices of $\bU_{v,[b-V+1]}$ and $\bhY_{v,[b-1]}$, respectively. Define $\mathbf{S}_b=(S_{\ml_1[b]},\cdots,S_{\ml_{\ell}[b-\ell+1]})$. Also, let $\mathbf{s}=(s_{\ml_1},\cdots,s_{\ml_{\ell}})$, in which $s_v=(w_v,z_v):w_v\in[1,2^{n(H(U_v)+\delta)}],z_v\in[1,2^{n(I(Y_v;\hY_v|X_v)+\delta)}]$. Define the events,
\begin{align}
\me_0(b,k)&:=\{(\bU_{\ml_k,[b-k-V+2]},\bU_{\ml^k,[b-k-V+2]},\bU_{d_i,[b-k-V+2]})\notin\styp\} \nonumber\\
\me_1(b,k,v)&:=\{(\bX_{v,[b-k+1]},\bY_{v,[b-k+1]},\bhY_{v}(z_v|\bX_{v,[b-k+1]}))\notin\stypo,\ \mbox{for all $z\in[1,2^{n(I(Y_v;\hY_v)+\delta)}]$}\}\n
\me_2(b,k,\mathbf{s})&:=\{(\uu_{\ml_k}(w_{\ml_k}),\bU_{\ml^k,[b-k-V+2]},\bU_{d_i,[b-k-V+2]})\in\styp\}\n
\me_3(b,k,\mathbf{s})&:=\{(\bX_{\ml_k}(s_{\ml_k}),\bhY_{\ml_{k-1}}(z_{\ml_{k-1}}|\bX_{\ml_{k-1}[b-k+1]}),\bX_{\ml^k,[b-k+1]},
  \bhY_{\ml^{k-1},[b-k+1]},\bY_{d_i,[b-k+1]},\bX_{d_i,[b-k+1]})\in\styp\}.
 \end{align}
 Then the error event $\me(b)$ corresponding to decoding at the end of block $b$ can be expressed as 
 \[ \me(b)=\cup_{k=1}^{\ell+1}\big(\me_0(b,k)\bigcup\cup_{v\in\mv}\me_1(b,k,v)\bigcup\me_3^C(b,k,\mathbf{S}_b)\big)\bigcup\cup_{\mathbf{s}\neq\mathbf{S}_b}\big(\cap_{k=1}^{\ell+1}\me_2(b,k,\mathbf{s})\cap\me_3(b,k,\mathbf{s})\big).\]
 Using the union bound, we bound above the probability of error as follows: 
\begin{align}
\mathbb{P}[\me(b)]&\le \mathbb{P}[\cup_{k=1}^{\ell+1}\me_0(b,k)]+\mathbb{P}[\cup_{k=1}^{\ell+1}\cup_{v\in\mv}\me_1(b,k,v)]+\mathbb{P}[\cup_{k=1}^{\ell+1}(\me_3^C(b,k,\mathbf{S}_b)\bigcap\cap_{v\in\mv}\me_1^C(b,k,v))]+\n
&\qquad \mathbb{P}[\cup_{\mathbf{s}\neq\mathbf{S}_b}\big(\cap_{k=1}^{\ell+1}\me_2(b,k,\mathbf{s})\cap\me_3(b,k,\mathbf{s})\big)].
\end{align}
By the typical lemma \cite[Theorem 1.1]{kramer:book}, the first term vanishes as $n\rightarrow\infty$, the second term vanishes since at each node $v$ and for each input $\ux_{v,[b-k+1]}$, there are more than $2^{nI(Y_v;\hY_v|X_v)}$ codewords $\uhy_v(z_v|\ux_{v,[b-k+1]})$, and the third term vanishes by \cite[Theorem 1.2.]{kramer:book}. For the last term, let $\me_1(b)=\cup_{\mathbf{s}\neq\mathbf{S}_b}\big(\cap_{k=1}^{\ell+1}\me_2(b,k,\mathbf{s})\cap\me_3(b,k,\mathbf{s})\big)$.
\begin{align}
\mathbb{P}[\me_1(b)]&\le\sum_{\mathclap{\uu_{\ma[b-\ell-V+2]},\cdots,\uu_{\ma[b-V+1]},\mathbf{s}_b}}p(\uu_{\ma[b-\ell-V+2]})\cdots p(\uu_{\ma[b-V+1]})\mathbb{P}[\mathbf{S}_b=\mathbf{s}_b|\uu_{\ma[b-\ell-V+2]},\cdots,\uu_{\ma[b-V+1]}]\n
&\qquad\sum_{\mathbf{s}\neq\mathbf{s}_b}\mathbb{P}[\cap_{k=1}^{\ell+1}\me_2(b,k,\mathbf{s})|\uu_{\ma[b-\ell-V+2]},\cdots,\uu_{\ma[b-V+1]}]\mathbb{P}[\cap_{k=1}^{\ell+1}\me_3(b,k,\mathbf{s})|\mathbf{S}_b=\mathbf{s}_b]\label{sal:1}\\
&=\sum_{\mathclap{\uu_{\ma[b-\ell-V+2]},\cdots,\uu_{\ma[b-V+1]},\mathbf{s}_b}}p(\uu_{\ma[b-\ell-V+2]})\cdots p(\uu_{\ma[b-V+1]})\mathbb{P}[\mathbf{S}_b=\mathbf{s}_b|\uu_{\ma[b-\ell-V+2]},\cdots,\uu_{\ma[b-V+1]}]\n
&\qquad\sum_{\mathbf{s}\neq\mathbf{s}_b}\prod_{k=1}^{\ell+1}
\mathbb{P}[\me_2(b,k,\mathbf{s})|\uu_{\ma[b-k-V+2]}]\mathbb{P}[\me_3(b,k,\mathbf{s})|\mathbf{S}_b=\mathbf{s}_b]\label{sal:2}\\
&=\sum_{\mathclap{\uu_{\ma[b-\ell-V+2]},\cdots,\uu_{\ma[b-V+1]},\mathbf{s}_b}}p(\uu_{\ma[b-\ell-V+2]})\cdots p(\uu_{\ma[b-V+1]})\mathbb{P}[\mathbf{S}_b=\mathbf{s}_b|\uu_{\ma[b-\ell-V+2]},\cdots,\uu_{\ma[b-V+1]}]\n
&\qquad\sum_{\mathbf{s}\neq\mathbf{s}_b}\prod_{k=1}^{\ell+1}
\mathbf{1}[(\uu_{\ml_k}(w_{\ml_k}),\uu_{\ml^k,[b-k-V+2]},\uu_{d_i,[b-k-V+2]})\in\styp]\mathbb{P}[\me_3(b,k,\mathbf{s})|\mathbf{S}_b=\mathbf{s}_b]\label{sal:3}
\end{align} 
where \eqref{sal:1} follows from the fact that the codebook generation is independent of the sources $U_{\ma}^{nB}$, \eqref{sal:2} follows from the fact that the codebooks used in any $\ell\le V$ consecutive blocks are generated independently and  the fact that the sources are i.i.d., therefore the source sequences are independently generated  in the consecutive blocks and \eqref{sal:3} follows from the definition of $\me_2(b,k,\mathbf{s})$, in which $\mathbf{1}$ represents the indicator function. Define,
\be\label{sal:def-mn}\begin{split}
 \mn_{\ms,\mz}(\uu_{\ma[b-\ell-V+2]},\cdots,\uu_{\ma[b-V+1]},\mathbf{s}_b)=\Big\{\mathbf{s}: s_t\neq s_{t,[b-k+1]}, z_{t'}\neq z_{t',[b-k]},\\ \mbox{for all}\ k\in[1,\ell+1],t\in\ms\cap\ml_k,t'\in\mz\cap\ml_k,    
  \ \mbox{and}\ s_{\ml_k\backslash\ms}=s_{\ml_k\backslash\ms,[b-k+1]}, \mbox{for all $k\in[1,\ell]$}, \\
 \mbox{and}\quad  (\uu_{\ml_k}(w_{\ml_k}),\uu_{\ml^k,[b-k-V+2]},\uu_{d_i,[b-k-V+2]})\in\styp,\ \mbox{for all $k\in[1,\ell]$}\Big\}.
\end{split}\ee
Then, \eqref{sal:3} can be rewritten as,
\begin{align}
\mathbb{P}[\me_1(b)]&\le\sum_{\mathclap{\uu_{\ma[b-\ell-V+2]},\cdots,\uu_{\ma[b-V+1]},\mathbf{s}_b}}p(\uu_{\ma[b-\ell-V+2]})\cdots p(\uu_{\ma[b-V+1]})\mathbb{P}[\mathbf{S}_b=\mathbf{s}_b|\uu_{\ma[b-\ell-V+2]},\cdots,\uu_{\ma[b-V+1]}]\n
&\qquad
\sum_{\emptyset\neq\ms\subseteq\mv_{-d_i}}\sum_{\mz\subseteq\ms}\sum_{\mathbf{s}\in \mn_{\ms,\mz}(\uu_{\ma[b-\ell-V+2]},\cdots,\uu_{\ma[b-V+1]},\mathbf{s}_b)}\prod_{k=1}^{\ell+1}\mathbb{P}[\me_3(b,k,\mathbf{s})|\mathbf{S}_b=\mathbf{s}_b]\label{sal:5}
\end{align} 
Define,
\be\label{sal:le}
\mathbb{P}_{\ms,\mz}=\sum_{\mathbf{s}\in \mn_{\ms,\mz}(\uu_{\ma[b-\ell-V+2]},\cdots,\uu_{\ma[b-V+1]},\mathbf{s}_b)}\prod_{k=1}^{\ell+1}\mathbb{P}[\me_3(b,k,\mathbf{s})|\mathbf{S}_b=\mathbf{s}_b].\ee
Notice that there are $3^{|\mv_{-d_i}|}$ pairs $(\ms,\mz)$ such that $\mz\subseteq\ms\subseteq\mv_{-d_i}$. Using this fact, $\mathbb{P}[\me_1(b)]$ is upper bounded by,
\be
\mathbb{P}[\me_1(b)]\le 3^{|\mv_{-d_i}|}\max_{\mz\subseteq\ms\subseteq\mv_{-d_i}}\mathbb{P}_{\ms,\mz}.
\ee
Therefore to show that $\mathbb{P}[\me_1(b)]$ vanishes as $n\rightarrow\infty$, it suffices to show that each of $\mathbb{P}_{\ms,\mz}$ vanishes as $n\rightarrow\infty$. To bound above the probability $\mathbb{P}_{\ms,\mz}$, we use the following lemmas which provide an upper bound on the probability inside the last summation.
\begin{lemma}\label{le:9}
For each $\mathbf{s}\in \mn_{\ms,\mz}(\uu_{\ma[b-\ell-V+2]},\cdots,\uu_{\ma[b-V+1]},\mathbf{s}_b)$, we have
\bes
\mathbb{P}[\me_3(b,k,\mathbf{s})|\mathbf{S}_b=\mathbf{s}_b]\stackrel{.}{\le} 2^{-n\beta_{\ms,\mz}(k)},
\ees
where $\beta_{\ms,\mz}(k)$ is given by
\be\label{sal:simp1}
\beta_{\ms,\mz}(k)=\sum_{t\in\ms\cap\ml_k}H(X_t)+\sum_{t'\in\mz\cap\ml_{k-1}}H(\hY_{t'}|X_{t'})-H(X_{\ms\cap\ml_k}\hY_{\mz\cap\ml_{k-1}}|X_{\ms^C\cap\ml_k}\hY_{\mz^C\cap\ml_{k-1}}X_{\ml^k}\hY_{\ml^{k-1}}X_{d_i}Y_{d_i}).
\ee
\end{lemma}
\begin{proof}
See Appendix \ref{app:5}.
\end{proof}
\begin{lemma}\label{le:7}
For each $(\uu_{\ma[b-\ell-V+2]},\cdots,\uu_{\ma[b-V+1]},\mathbf{s}_b)$, we have
\bes
\mn_{\ms,\mz}(\uu_{\ma[b-\ell-V+2]},\cdots,\uu_{\ma[b-V+1]},\mathbf{s}_b)\stackrel{.}{\le} 2^{n(\sum_{k=1}^{\ell}H(U_{\ms\cap\ml_k}|U_{\ml_k\backslash\ms}U_{\ml^k}U_{d_i})+\sum_{t\in\mz}I(Y_t;\hY_t|X_t))}.
\ees
\end{lemma}
\begin{proof}
See Appendix \ref{app:6}.
\end{proof}
Applying Lemma \ref{le:9} and Lemma \ref{le:7} to \eqref{sal:le} yields,
\be 
\mathbb{P}_{\ms,\mz}\stackrel{.}{\le} 2^{-n(\sum_{k=1}^{\ell+1}\beta_{\ms,\mz}(k)-\sum_{k=1}^{\ell}H(U_{\ms\cap\ml_k}|U_{\ml_k\backslash\ms}U_{\ml^k}U_{d_i})-\sum_{t'\in\mz}I(Y_{t'};\hY_{t'}|X_{t'}))}.
\ee
Thus $\mathbb{P}_{\ms,\mz}$ vanishes as $n\rightarrow\infty$, provided that 
\be\label{sal:simp2}
\sum_{k=1}^{\ell+1}\beta_{\ms,\mz}(k)-\sum_{k=1}^{\ell}H(U_{\ms\cap\ml_k}|U_{\ml_k\backslash\ms}U_{\ml^k}U_{d_i})-\sum_{t'\in\mz}I(Y_{t'};\hY_{t'}|X_{t'})> 0.
\ee
Substituting \eqref{sal:simp1} in \eqref{sal:simp2}, simplifies it as follows,
\begin{align}
0&<\sum_{t\in\ms}H(X_t)+\sum_{t'\in\mz}H(\hY_t|X_tY_t)-\sum_{k=1}^{\ell+1}\left(H(X_{\ms\cap\ml_k}\hY_{\mz\cap\ml_{k-1}}|X_{\ml_k\backslash\ms}\hY_{\ml_{k-1}\backslash\mz}X_{\ml^{k}}\hY_{\ml^{k-1}}X_{d_i}Y_{d_i})\right.\n
&\qquad\qquad\qquad\qquad\qquad\qquad\qquad\qquad\left.+H(U_{\ms\cap\ml_k}|U_{\ml_k\backslash\ms}U_{\ml^k}U_{d_i})\right)\label{sal:simp}\\
&=\sum_{k=1}^{\ell+1}\left(H(X_{\ms\cap\ml_k})+H(\hY_{\mz\cap\ml_{k-1}}|X_{\mz\cap\ml_{k-1}}Y_{\mz\cap\ml_{k-1}})-H(X_{\ms\cap\ml_k}\hY_{\mz\cap\ml_{k-1}}|X_{\ml_k\backslash\ms}\hY_{\ml_{k-1}\backslash\mz}X_{\ml^{k}}\hY_{\ml^{k-1}}Y_{d_i})\right.\n
&\qquad\qquad\qquad\left.-H(U_{\ms\cap\ml_k}|U_{\ml_k\backslash\ms}U_{\ml^k}U_{d_i})\right)\label{sal:7}\\
&= \sum_{k=1}^{\ell+1}\left(I(X_{\ms\cap\ml_k};Y_{d_i}\hY_{(\ml_{k-1}\backslash\mz)\cup\ml^{k-1}}|X_{d_i}X_{(\ml_k\backslash\ms)\cup \ml^k})-I(Y_{\mz\cap\ml_{k-1}};\hY_{\mz\cap\ml_{k-1}}|X_{d_i}X_{\ml^{k+1}}Y_{d_i}\hY_{(\ml_{k-1}\backslash\mz)\cup\ml^{k-1}})\right.\n
&\qquad\qquad\left. -H(U_{\ms\cap\ml_k}|U_{\ml_k\backslash\ms}U_{\ml^k}U_{d_i})\right)\label{sal:8}
\end{align}
where in \eqref{sal:7} and \eqref{sal:8}, we have used the fact that $X_t$'s are independent and $\hY_t$ given $(X_t,Y_t)$ is independent of all the other random variables. Now consider the RHS of \eqref{sal:8}. Since $\mz\subseteq\ms$, it can easily be shown that the first term inside the summation takes its minimum for $\mz=\ms$ while the second term simultaneously takes its maximum for $\mz=\ms$. On the other side, $\mz=\ms$ corresponds to the probability $\mathbb{P}_{\ms,\ms}$. Hence if $\mathbb{P}_{\ms,\ms}$ vanishes, then all $\mathbb{P}_{\ms,\mz}:\mz\subseteq\ms$ vanish as $n\rightarrow\infty$. Therefore $\mathbb{P}[\me_1(b)]$ vanishes if all  $\mathbb{P}_{\ms,\ms}$ ($\ms\subseteq\mv_{-d_i}$) vanish. Finally, substituting $\mz=\ms$ in \eqref{sal:simp} results in \eqref{eq:sufficient}, which completes the proof of Lemma \ref{le:first}. \par
\begin{remark}
If there exists only a single destination, one can use the offset encoding scheme of \cite{xie} and \cite{kramer2005} which has less delay compared to the proposed encoding scheme, to prove Lemma \ref{le:first}. In general however, since the ordered partitions corresponding to each receiver for reliable decoding are different, it is impossible to obtain the same offset encoding scheme for all the destinations. This makes it clear why the encoding scheme does not transmit any information in the first $V-1$ blocks.  
\end{remark}     
\subsection{Removing additional constraints}
\label{sub:c}
This subsection claims that for each $d_i$, we can reduce the constraints of \eqref{eq:adi} to the first term of it. A special case of our claim about the single relay channel has been studied in \cite{kang:itw}. We prove our claim by induction on $|\mv_{-d_i}|$. For $|\mv_{-d_i}|=1$, it is true. Now suppose the induction assumption is true for all $k<|\mv_{-d_i}|$. 
For each $\mz\subseteq\mv$ which contains $d_i$ and each $\ms\subseteq\mz\backslash\{d_i\}$, let
\[h^{(d_i)}_{\mz}(\ms)=R^{(d_i)}_{\ms} - H(\hY_{\ms}X_{\ms}|X_{\mz\backslash\ms}\hY_{\mz\backslash(\ms\cup\{d_i\})}Y_{d_i})\] 
\par Assume there exists a subset $\mt$ of $\ma^C\backslash\{d_i\}$ such that $h^{(d_i)}_{\mv}(\mt)<0$. For each $\mw\subseteq\mv_{-d_i}$ observe that,
\begin{IEEEeqnarray}{rLl}
h^{(d_i)}_{\mv}(\mw\cup\mt)&=&h^{(d_i)}_{\mv}(\mt)+R^{(d_i)}_{\mw\backslash\mt}-
 H(\hY_{\mw}X_{\mw}|X_{\mw^C\backslash\mt}\hY_{\mw^C\backslash(\mt\cup\{d_i\})}Y_{d_i}) \nonumber\\
&<&R^{(d_i)}_{\mw\backslash\mt}-H(\hY_{\mw}X_{\mw}|X_{\mw^C\backslash\mt}\hY_{\mw^C\backslash(\mt\cup\{d_i\})}Y_{d_i}) \nonumber\\
&\le&R^{(d_i)}_{\mw}-H(\hY_{\mw}X_{\mw}|X_{\mw^C}\hY_{\mw^C\backslash\{d_i\}}Y_{d_i}) \nonumber\\  \label{eq:symp}
&=&h^{(d_i)}_{\mv}(\mw)
\end{IEEEeqnarray}
Using \eqref{eq:symp}, \eqref{eq:adi1} can be simplified as follows:
\begin{IEEEeqnarray}{rCl}
H(U_{\ms}|U_{\ma\backslash\ms})&\le& \min_{\mv\supset\mw\supseteq\ms:\atop d_i\in\mw^C} h^{(d_i)}_{\mv}(\mw)\nonumber\\
&\stackrel{(a)}{=}&\min_{\mv\supset\mw\supseteq\ms:\atop d_i\in\mw^C} h^{(d_i)}_{\mv}(\mw\cup\mt)\nonumber\\
&\stackrel{(b)}{\le}&\min_{\mv\supset\mw\supseteq\ms:\atop d_i\in\mw^C}h^{(d_i)}_{\mv\backslash\mt}(\mw\backslash\mt)\nonumber\\
\label{eq:fin}
&=& \min_{\mv\backslash\mt\supset\mw\supseteq\ms:\atop d_i\in\mw^C}h^{(d_i)}_{\mv\backslash\mt}(\mw)
\end{IEEEeqnarray}  
where (a) follows from \eqref{eq:symp}, since $\ms\subset\mw\cup\mt$ and $d_i\notin\mt$, and (b) follows from the first inequality in \eqref{eq:symp}.\par
Now by the induction assumption, the last term of \eqref{eq:fin} corresponds to the feasibility constraints of the reliable transmission of $U_{\ma}$ to node $d_i$ over the cooperative network with the set of nodes $\mv\backslash\mt$. Hence node $d_i$ can decode $U_{\ma}$, by treating $(X_{\mt},\hY_{\mt})$ as noise. We note that the encoding scheme, only results in more delay rather than a corresponding encoding/decoding scheme for a cooperative networks with the node's set $\mv\backslash\mt$. Therefore, the encoding scheme does not need any changes and the decoding is done only with respect to the cooperative network with $\mv=\mv\backslash\mt$. This proves our claim.

\section{Slepian-Wolf coding over some classes of cooperative networks}\label{sec:6}
In this section, we extract some corollaries from Proposition \ref{pro:ob} and Theorem \ref{thm:sw} about semi-deterministic network, Aref networks and linear finite-field and state-dependent deterministic networks, for which Proposition \ref{pro:ob} and Theorem \ref{thm:sw} (partially) match.   
\begin{definition}
A cooperative network with one destination $d$, is said to be \emph{semi-deterministic}, if each node $v\in\mv\backslash\{d\}$ observes a deterministic function of all the channel inputs and the destination channel output, i.e., $Y_v=f_v(X_{\mv},Y_d)$.
\end{definition}
\begin{remark}
The semi-deterministic cooperative network is a generalization of semi-deterministic relay channel \cite{aref} and a class of deterministic relay channels, recently defined in \cite{yhk}.
\end{remark}
\begin{definition}
A cooperative network is said to be \emph{deterministic}, if each node observes a deterministic function of all the channel inputs, i.e., $Y_v=f_v(X_{\mv})$.
\end{definition}
\begin{definition}
A deterministic network is said to be an \emph{Aref network}, if each channel output $Y_v$ can be decomposed into $|\mv|-1$ components $(Y_{v',v}:v'\in\mv\backslash\{v\})$, where $Y_{v',v}$ is a deterministic function of $X_{v'}$. A semi-deterministic network with destination node $d$, is said to be a \emph{semi-deterministic Aref network}, if each channel output $Y_v$ can be decomposed into $|\mv|-1$ components $(Y_{v',v}:v'\in\mv\backslash\{v\})$, where $Y_{v',v}$ is a deterministic function of $X_{v'}$ for $v\in\mv_{-d}$ and $Y_{v',d}$ is a stochastic function of $X_{v'}$.
\end{definition}
\begin{definition}
A deterministic network is said to be a \emph{linear finite-field deterministic network}, if all the channel inputs and outputs lie in the same field $\mathbf{GF}(q)$ and each channel output can be expressed as a linear combination of all the channel inputs. The relation between the channel inputs and the channel outputs can be determined via a matrix product, $Y_{\mv}=\mathbf{G}X_{\mv}$, where $\mathbf{G}$ is called the channel matrix of the network. $\mathbf{G}_{\mt_1,\mt_1}$ is a sub-matrix obtained by deleting the rows and columns of $\mathbf{G}$ corresponding to $\mt_1$ and $\mt_2$, respectively.
\end{definition}
\begin{definition}
A cooperative network is state-dependent (SD) \cite{yhk:isit09}, if there exists a set of states $\ms$ such that the channel inputs and the channel outputs at each time are related via the current state of the network. A SD-cooperative network is said to be deterministic if each node observes a deterministic function of all the channel inputs and the state of the network, i.e., $Y_v=f_v(X_{\mv},S)$. A SD-deterministic network is said to be an Aref network, if each channel output $Y_v$ can be decomposed into $|\mv|-1$ components $(Y_{v',v}:v'\in\mv\backslash\{v\})$, where $Y_{v',v}$ is a deterministic function of $(X_{v'},S)$. A SD-linear finite-field deterministic network is a network described by $Y_{\mv}=\mathbf{G}(S)X_{\mv}$, where $\mathbf{G}(S)$ is the matrix of coefficients corresponding to state $S$.
\end{definition}
\begin{proposition}
\label{pro:semi}
The set of DMCS $U_{\ma}$ can reliably be transmitted over a semi-deterministic network, if there exists random variable $Q$, such that for each $\ms\subseteq\mv$, we have :
\begin{equation}
\label{eq:semi}
H(U_{\ms}|U_{\ma\backslash\ms})<\min_{\mv_{-d}\supseteq\mw\supseteq\ms}I(X_{\mw};Y_{\mw^C}|X_{\mw^C}Q)
\end{equation}
where the joint p.m.f. of random variables factors as $p(q)[\prod_{v\in\mv}p(x_v|q)]p(y_{\mv}|x_{\mv})$.\\
On the other side, multicasting is feasible, only if there 
exists a joint p.m.f. $p(x_{\mv})$ such that 
\begin{equation}
H(U_{\ms}|U_{\ma\backslash\ms})<\min_{\mv_{-d}\supseteq\mw\supseteq\ms}I(X_{\mw};Y_{\mw^C}|X_{\mw^C}).
\end{equation}
\end{proposition}
\begin{proposition}
\label{pro:det}
The set of DMCS $U_{\ma}$ can reliably be multicast over a deterministic network, if there exists a product distribution $\prod_{v\in\mv}p(x_v)$ such that for each $\ms\subseteq\mv$, we have:
\begin{equation}
\label{eq:det}
H(U_{\ms}|U_{\ma\backslash\ms})<\min_{d_i\in\md}\min_{\mv_{-d_i}\supseteq\mw\supseteq\ms}H(Y_{\mw^C}|X_{\mw^C})
\end{equation}
On the other side, multicasting is feasible, only if there 
exists a joint p.m.f. $p(x_{\mv})$ such that 
\begin{equation}
H(U_{\ms}|U_{\ma\backslash\ms})<\min_{d_i\in\md}\min_{\mv_{-d_i}\supseteq\mw\supseteq\ms}H(Y_{\mw^C}|X_{\mw^C}).
\end{equation}
\end{proposition} 
\begin{remark}
Comparing the direct part and converse part of Propositions \ref{pro:semi} and \ref{pro:det}, we see that the sufficient conditions partially match to necessary conditions and these conditions completely match together, if  we can restrict the set of joint p.m.f. in the converse part to the set of product distributions.
  \end{remark}
\begin{proposition}
\label{pro:aref}
The set of DMCS $U_{\ma}$ can reliably be multicast over an Aref network, if and only if, there exists a product distribution $\prod_{v\in\mv}p(x_v)$ such that for each $\ms\subseteq\mv$, we have:
\begin{equation}
\label{eq:aref}
H(U_{\ms}|U_{\ma\backslash\ms})<\min_{d_i\in\md}\min_{\mv_{-d_i}\supseteq\mw\supseteq\ms}\sum_{v\in\mw}H(Y_{v,\mw^C})
\end{equation}
\end{proposition}
\begin{remark}
This proposition was partially proved in \cite{babu} for acyclic Aref networks.
\end{remark} 
\begin{proposition}
\label{pro:semi-aref}
The set of DMCS $U_{\ma}$ can reliably be transmitted over a semi-deterministic Aref network, if and only if, there exists a product distribution $\prod_{v\in\mv}p(x_v)$ such that for each $\ms\subseteq\mv$, we have:
\begin{equation}
\label{eq:semi-aref}
H(U_{\ms}|U_{\ma\backslash\ms})<\min_{\mv_{-d}\supseteq\mw\supseteq\ms}\sum_{v\in\mw}I(X_v;Y_{v,\mw^C}) 
\end{equation}
\end{proposition} 
\begin{proposition}
\label{pro:finite}
The set of DMCS $U_{\ma}$ can reliably be multicast over a linear finite-field deterministic network, if and only if, 
\begin{equation}
\label{eq:finite}
H(U_{\ms}|U_{\ma\backslash\ms})<\min_{d_i\in\md}\min_{\mv_{-d_i}\supseteq\mw\supseteq\ms}\mbox{rank}(\mathbf{G}_{\mw,\mw^C})\log q 
\end{equation}
\end{proposition}
Now, consider the SD-network. In the sequel, assume that the state $S$ is an i.i.d. random process.   
\begin{proposition}
\label{pro:state}
For reliable multicasting over a SD-deterministic network, if all destinations have the state information $S$, then a sufficient condition is given by,
\begin{equation}
\label{eq:state}
\forall\ms\subseteq\ma: H(U_{\ms}|U_{\ma\backslash\ms})<\min_{d_i\in\md}\min_{\mv_{-d_i}\supseteq\mw\supseteq\ms}H(Y_{\mw^C}|X_{\mw^C},S)
\end{equation}
Moreover, condition \eqref{eq:state} is a necessary condition for reliable multicasting over a SD-Aref network and a SD-linear finite-field deterministic network with state information available at the destinations. In these cases, \eqref{eq:state} is simplified to,
\begin{align*}
\mbox{SD-Aref network}:& H(U_{\ms}|U_{\ma\backslash\ms})<\min_{d_i\in\md}\min_{\mv_{-d_i}\supseteq\mw\supseteq\ms}\sum_{v\in\mw}H(Y_{v,\mw^C}|S)\\
\mbox{SD-linear finite-field deterministic network}:&
H(U_{\ms}|U_{\ma\backslash\ms})<\min_{d_i\in\md}\min_{\mv_{-d_i}\supseteq\mw\supseteq\ms}\mathbb{E}_{S}[\mbox{rank}(\mathbf{G}_{\mw,\mw^C}(S))] \log q 
\end{align*} 
\end{proposition} 
\begin{proof}[Proof of Propositions 2-7]
The direct part of Propositions \ref{pro:semi} and \ref{pro:det} follow from Theorem \ref{thm:sw}, by setting $\hY_v=Y_v$ in Theorem \ref{thm:sw}, because $(Y_{\mw}: \mw\subseteq\mv_{-d})$ and $(Y_{\mw}: \mw\subseteq\mv)$ are deterministic functions of $(Y_{d},X_{\mv})$ and $X_{\mv}$, respectively. The converse part of Propositions \ref{pro:semi} and \ref{pro:det} are the direct consequence of Proposition \ref{pro:ob}. The direct part of Proposition \ref{pro:aref} follows from Proposition \ref{pro:det}, and the converse is deduced from Proposition \ref{pro:ob} as follows:
\begin{align*}
H(U_{\ms}|U_{\ma\backslash\ms})&<H(Y_{\mw^C}|X_{\mw^C})\\
&\le H(\cup_{v\in\mw}Y_{v,\mw^C})\\
&\le \sum_{v\in\mw}H(Y_{v,\mw})
\end{align*} 
Now, since $Y_{v,\mw^C}$ depends only on $X_v$, the last term of inequalities only depends on the mariginal p.m.f. of the random variables. Thus, we can restrict the set of joint p.m.f. of Proposition \ref{pro:ob} to the product distribution, which completes the proof of Proposition \ref{pro:aref}. The direct part of Proposition \ref{pro:semi-aref} follows from Proposition \ref{pro:semi} and the converse part is obtained from Proposition \ref{pro:ob} as follows:
\begin{align}
H(U_{\ms}|U_{\ma\backslash\ms})&<I(X_{\mw};Y_{\mw,\mw^C}Y_{\mw^C,\mw^C}|X_{\mw^C})\label{eq:semi-aref-1}\\
&=I(X_{\mw};Y_{\mw,\mw^C}|X_{\mw^C}Y_{\mw^C,\mw^C})\label{eq:semi-aref-2}\\
&\le I(X_{\mw};Y_{\mw,\mw^C})\label{eq:semi-aref-3}\\
&\le \sum_{v\in\mw}I(X_v;Y_{v,\mw^C})\label{eq:semi-aref-4} 
\end{align}
where \eqref{eq:semi-aref-2} follows, because $X_{\mw}-X_{\mw^C}-Y_{\mw^C,\mw^C}$ form a Markov chain and \eqref{eq:semi-aref-3} follows from the fact that $(X_{\mw^C}Y_{\mw^C,\mw^C})-X_{\mw}-Y_{\mw,\mw^C}$ form a Markov chain and \eqref{eq:semi-aref-4} follows, since $Y_{v,\mw^C}$ given $X_v$ is independent of other random variables. Finally, note that the RHS of \eqref{eq:semi-aref-4} only depends on the marginal p.m.f. of the random variables $X_{\mv}$ which implies the converse.\par
The direct part of Proposition \ref{pro:finite} is deduced from Proposition \ref{pro:det}, by computing the RHS of \eqref{eq:det} for the product distribution $\prod_{v\in\mv}p(x_{v})$, in which each $X_v$ is uniformly distributed  over the field $\mathbf{GF}(q)$. The converse follows from Proposition \ref{pro:ob}, since the product and the uniform distribution simultaneously maximized the RHS of \eqref{eq:sw2} for all $\mw\subseteq\mv$.\par
The sufficient condition of Proposition \ref{pro:state} is deduced from Theorem \ref{thm:sw}, by treating the state information at each destination as an additional output of the network and the fact that $(Y_{\mw}:\mw\subseteq\mv_{-d_i})$ is a deterministic function of $(X_{\mv},S)$. The necessary conditions for the SD-Aref network and the SD-linear finite-field deterministic network follow from similar arguments for the converse of these networks without state. 
\end{proof} 
\section{Slepian-Wolf coding over Gaussian cooperative networks}\label{sec:7}
In the previous section, we focused on some networks for which the cut-set type necessary conditions became sufficient conditions at least for product distribution of channel inputs. In this section, we focus on the Gaussian networks for which simple forwarding of the observations of each node is impossible. Instead, following \cite{aves:phd,aves:sub}, each node quantizes its observations at the noise level, then transmits these to the destinations. We compute sufficient conditions corresponding to this approach and compare it with the necessary conditions. \par
Consider a Gaussian cooperative network, in which the received signal $\uy_{v}$ is given by,
\be
\uy_{v}=\sum_{v'\in\mv_{-v}}h_{v',v}\ux_{v'}+\uz_{v}
\ee  
\noindent where $h_{v',v}$ is a complex number which represents the channel gain from node $v'$ to node $v$. Furthermore, we assume that each node has an average power constraint equal to one on its transmitted signal. Moreover, $Z_v$ is an i.i.d. complex Gaussian random process with variance $\sigma^2_v$. Theorem \ref{thm:gauss} is the main result of this section.
\begin{theorem}\label{thm:gauss}
A set of DMCS $U_{\ma}$ can reliably be transmitted over a Gaussian network, if for each $\ms\subseteq\mv$, we have:
\begin{equation}
\label{eq:gauss-in}
H(U_{\ms}|U_{\ma\backslash\ms})<\min_{d_i\in\md}\min_{\mv_{-d_i}\supseteq\mw\supseteq\ms}C_{wf}(\mw\rightarrow\mw^C)-\kappa_{\mw}
\end{equation}
where 
\[
C_{wf}(\mw\rightarrow\mw^C)=\max_{p(x_{\mw}):\sum_{v\in\mw}\mathbb{E}X^2_v=|\mw|}I(X_{\mw};Y_{\mw^C}|X_{\mw^C})
\]
and 
\[
\kappa_{\mw}=\min\{|\mw|,|\mw^C|\}\log(1+\dfrac{|\mw|}{\min\{|\mw|,|\mw^C|\}})+V-1
\]
Moreover, $\kappa_{\mw}$ is bounded above by $\frac{3}{2}V-1$.\\
On the other side, the multicasting is feasible, only if:
  \begin{equation}
\label{eq:gauss-out}
H(U_{\ms}|U_{\ma\backslash\ms})<\min_{d_i\in\md}\min_{\mv_{-d_i}\supseteq\mw\supseteq\ms}C_{wf}(\mw\rightarrow\mw^C)
\end{equation} 
\end{theorem}
\begin{remark}
This theorem establishes the fact that multicasting of all DMCS whose Slepian-Wolf region intersects cut-set bound region within a $\dfrac{3}{2}V-1$ bits, is feasible.
\end{remark}
\begin{proof}
$C_{wf}(\mw\rightarrow\mw^C)$ is the capacity of the $\mw\times\mw^C$ MIMO channel with antenna input $X_{\mw}$ and antenna output $Y_{\mw^C}$. Now constraint \eqref{eq:gauss-out} is a direct result of Proposition \ref{pro:ob}, since there exists an average power constraint equal to one at each node $v\in\mw$. To show \eqref{eq:gauss-in}, we apply Theorem \ref{thm:sw} to the Gaussian network. Assume $(X_v:v\in\mw)$ be jointly complex Gaussian random variables with covariance matrix $I_{V\times V}$. Let $\hY_v=Y_v+\hat{Z}_v$ where $\hat{Z}_v$ is a complex Gaussian random variable with variance equal to $\sigma_v^2$ (In other words, $\hY_{v}$ quantizes $Y_v$ at the noise level, \cite{aves:sub}). Now consider,
\begin{align}
I(X_{\mw};Y_{\mw^C}|X_{\mw^C})&=I(X_{\mw};Y_{\mw^C}\hY_{\mw^C\backslash\{d_i\}}|X_{\mw^C})\label{eq:g-1}\\
&=I(X_{\mw};Y_{d_i}\hY_{\mw^C\backslash\{d_i\}}|X_{\mw^C})+I(X_{\mw};Y_{\mw^C\backslash\{d_i\}}|X_{\mw^C}\hY_{\mw^C\backslash\{d_i\}})\label{eq:g-2}
\end{align}
where \eqref{eq:g-1} follows, since $X_{\mw}-(X_{\mw^C},Y_{\mw^C})-\hY_{\mw^C}$ form a Markov chain. Next consider,
\begin{align}
I(X_{\mw};Y_{\mw^C\backslash\{d_i\}}|X_{\mw^C}\hY_{\mw^C\backslash\{d_i\}})&=I(X_{\mw};\hZ_{\mw^C\backslash\{d_i\}}|X_{\mw^C},Y_{\mw^C\backslash\{d_i\}}+\hZ_{\mw^C\backslash\{d_i\}})\label{eq:g-3}\\
&\le h(\hZ_{\mw^C\backslash\{d_i\}})-h(\hZ_{\mw^C\backslash\{d_i\}}|Z_{\mw^C\backslash\{d_i\}}+\hZ_{\mw^C\backslash\{d_i\}})\label{eq:g-4}\\
&= I(\hZ_{\mw^C\backslash\{d_i\}};Z_{\mw^C\backslash\{d_i\}}+\hZ_{\mw^C\backslash\{d_i\}})\label{eq:g-5}\\
&= |\mw^C|-1\label{eq:g-6}
\end{align}
where \eqref{eq:g-3} follows from the definition of $\hY_{v}$, \eqref{eq:g-4} follows from the fact that conditioning does not increase entropy and the fact that conditioning on $(Y_{\mw^C}+\hZ_{\mw^C},X_{\mv})$ is equivalent to conditioning on $(Z_{\mw^C}+\hZ_{\mw^C},X_{\mv})$ and $(Z_{\mw^C},\hZ_{\mw^C})$ is independent of $X_{\mv}$. \eqref{eq:g-6} follows, because $\{(Z_v,\hZ_v):v\in\mw^C\}$ are independent and $Z_v$ and $\hZ_v$ are complex Gaussian r.v. with the same variance. 
In a similar way consider,
\begin{align}
I(Y_{\mw};\hY_{\mw}|X_{\mv}Y_{d_i}\hY_{\mw^C\backslash\{d_i\}})&=I(Z_{\mw};Z_{\mw}+\hZ_{\mw}|X_{\mv},Z_{d_i},Z_{\mw^C}+\hat{Z}_{\mw^C})\nonumber
\\&=I(Z_{\mw};Z_{\mw}+\hZ_{\mw})\nonumber\\
&=|\mw|\label{eq:g}
\end{align}
Next, we derive a slight modified version of Beam-Forming Lemma \cite[Appendix F]{aves:sub}. $C_{wf}(\mw\rightarrow\mw^C)$ with water-filling is given by
\[
C_{wf}(\mw\rightarrow\mw^C)=\sum_{i=1}^n \log(1+Q_{ii}\lambda_i)
\]
where $n=\min(|\mw|,|\mw^C|)$ and $\lambda_i$'s are the singular values of the channel matrix of the MIMO channel and $Q_{ii}$ is given by water-filling solution satisfying, $\sum_{i=1}^n Q_{ii}=|\mw|$. 
Following \cite[Appendix F, Equations 140-143]{aves:sub}, we obtain,
\begin{align}\label{eq:g-7}
C_{wf}(\mw\rightarrow\mw^C)-I(X_{\mw};Y_{\mw}|X_{\mw^C})&\le n\log(1+\dfrac{|\mw|}{n})\\
&\le n\log(\dfrac{V}{n})\label{eq:gg}
\end{align}  
Finally, comparing \eqref{eq:g-2}, \eqref{eq:g-6}, \eqref{eq:g} and \eqref{eq:g-7} we get,
\be
I(X_{\mw};Y_{d_i}\hY_{\mw^C\backslash\{d_i\}}|X_{\mw^C})\ge C_{wf}(\mw\rightarrow\mw^C)-\kappa_{\mw}
\ee
Substituting it in \eqref{eq:sw}, we conclude that the constraint \eqref{eq:gauss-in} is a sufficient condition. Now note that $n\in [1,\frac{V}{2})$. Define $f(x)=x\log(\dfrac{V}{x})$ on $[1,\frac{V}{2}]$. $f$ is a convex function and gets its maximum at the end point $\frac{V}{2}$. Hence the RHS of \eqref{eq:gg} is equal to or less than $\dfrac{V}{2}$ which results in $\kappa_{\mw}\le\frac{3}{2}V-1$. 
\end{proof}
\section{Achievable rate region for cooperative relay networks}\label{sec:8}
Consider $\ma=\mv$ and the sources $(U_v:v\in\mv)$ are statistically independent and uniformly distributed over the sets $\mm_v=\{1,2,\cdots,2^{R_v}\}$, thus $H(U_v)=R_v$. Substituting these values in Theorem \ref{thm:sw}, we find an achievable rate region which is based on the CF, for \emph{cooperative relay networks} with multicast demands. 
\begin{theorem}
\label{thm:ach}
 A V-tuple $(R_1,R_2,\cdots,R_V)$ is contained in the achievable rate region of a cooperative network with multicast demands at each node $d_i\in\md$, if for each $\ms\subseteq\mv$ the following constraint holds:
  \be
\label{eq:ach}
R_{\ms}< \min_{d_i\in\md\backslash\ms}\min_{\mv\supseteq\mw\supseteq\ms: \atop d_i\in\mw^C} \big[I(X_{\mw};Y_{d_i}\hY_{\mw^C\backslash\{d_i\}}|X_{\mw^C}Q)-I(Y_{\mw};\hY_{\mw}|X_{\mv}Y_{d_i}\hY_{\mw^C\backslash\{d_i\}}Q)\big]^+
 \ee
where $[x]^+=\max\{x,0\}$ and the joint p.m.f. of $(q,x_{\mv},y_{\mv},\hy_{\mv})$ factors as $p(q)\prod_{v\in\mv}p(x_v|q)p(\hy_v|x_v,y_v,q)]p(y_{\mv}|x_{\mv})$.
\end{theorem} 
\begin{proof}
  Let $\mt$ be the largest subset of $\mv$ such that the RHS of \eqref{eq:sw} is non-negative  subject to each $\ms\subseteq\mt$ (Note that if two subsets $\mt_1,\mt_2$ have this property, then $\mt_1\cup\mt_2$ also has this property, hence $\mt$ is unique.). Substituting $R_{\ms}=H(U_{\ms}|U_{\ms^C})$ in Theorem \ref{thm:sw} yields that $U_{\mt}$ can reliably be multicast, if \eqref{eq:ach} holds. Hence $(R_1,\cdots,R_V)$ is achievable (Note that $R_v=0$ for each node $v\in\mt^C$).
 \end{proof} 
\begin{corollary}
\label{cor:rel-1}
Consider a relay network with node $1$ as a transmitter which has no channel output, i.e., $Y_1=\emptyset$, $N-2$ relay nodes $\{2,\cdots,N-1\}$ and node $N$ as a destination which has no channel input, i.e., $X_N=\emptyset$. Substituting $R_2=\cdots=R_{N}=0$ in Theorem \ref{thm:ach} gives the following achievable rate ($R_{CF}$) for relay network.  
\be
\label{eq:ach:rel}
R_{CF}=\min_{\ms\subseteq\mv:\atop 1\in\ms,N\in\ms^C}\big[I(X_{\ms};\hY_{\ms^C\backslash\{V\}}Y_N|X_{\ms^C}Q)-\\I(Y_{\ms};\hY_{\ms}|X_{\mv}Y_V\hY_{\ms^C\backslash\{V\}}Q)\big]^+
\ee
\end{corollary}
\begin{remark}
For the single relay channel, the achievable rate is reduced to the CF rate with time-sharing as given in \cite{elgamal}.
\end{remark}
\begin{remark}
In \cite{yassaee}, we obtain an achievable rate based on CF, which subsumes the CF rate given in \cite{kramer2005}, when the partial decoding part of the CF strategy is relaxed. The CF rate in \cite[Theorem 3]{yassaee} is given by:
\be 
R^*_{CF}=I(X_1;Y_V\hY_{\mv_{-V}}|X_{\mv_{-V}})
\ee
\emph{subject to the constraints}  
\be
\label{eq:cf-cons}
\forall{\ms\subseteq\mv\backslash\{1,V\}}: I(Y_{\ms};\hY_{\ms}|X_{\mv_{-1}}Y_V\hY_{\ms^C\backslash\{V\}})\le I(X_{\ms};Y_V\hY_{\ms^C\backslash\{V\}}|X_{\ms^C\backslash\{V\}})
\ee
Now let $\mq=\emptyset$ in Corollary \ref{cor:rel-1}. It can be easily shown that when the constraints \eqref{eq:cf-cons} hold, then $\ms=\mv$ reaches the minimum of the RHS of \eqref{eq:ach:rel}. Therefore, the rate of Corollary \ref{cor:rel-1} subsumes the CF-rate given in \cite[Theorem 3]{yassaee}. 
\end{remark} 
\begin{corollary}
Consider a two-way relay network with nodes $1$ and $V$ as the two transmitters each demanding the message of the other one, and $V-2$ relay nodes $\{2,\cdots,V-1\}$. Substituting $R_2=\cdots=R_{V-1}=0$ and $\hY_1=\hY_V=\emptyset$ in Theorem \ref{thm:ach} gives the following achievable rate region for the two-way relay network.
\be
k=1,V:\  R_{k}=\min_{\ms\subseteq\mv:\atop k\in\ms,\bar{k}\in\ms^C}\big[I(X_{\ms};\hY_{\ms^C\backslash\{\bar{k}\}}Y_{\bar{k}}|X_{\ms^C})-\\I(Y_{\ms\backslash\{k\}};\hY_{\ms\backslash\{k\}}|X_{\mv}Y_{\bar{k}}\hY_{\ms^C\backslash\{\bar{k}\}})\big]^+
\ee
\noindent where $\bar{1}=V$ and $\bar{V}=1$.
\end{corollary}
\begin{remark}
Propositions \ref{pro:semi}-\ref{pro:state} are generalizations of several recent works on deterministic relay networks including \cite[Theorem 3.9]{aref}, \cite[Theorem 4.2]{aves:sub}, \cite[Theorem 4.4]{aves:sub}, \cite[Theorem 1]{yhk}, \cite[Theorem 1]{multicast} and \cite[Theorem 1]{yhk:isit09}.
\end{remark}\par
Next, consider the Gaussian cooperative network. Applying Theorem \ref{thm:gauss} to $U_{\mv}$, we conclude the following corollary which shows that the cut-set bound region is achievable within a constant number of bits. 
\begin{corollary}\label{cor:gauss}
A V-tuple $(R_1,R_2,\cdots,R_V)$ is contained in the achievable rate region of a Gaussian cooperative network with multicast demands at each node $d_i\in\md$, if for each $\ms\subseteq\mv$ the following constraint holds:
\begin{equation}
\label{eq:rel-gauss-in}
R_{\ms}<\min_{d_i\in\md}\min_{\mv_{-d_i}\supseteq\mw\supseteq\ms}C_{wf}(\mw\rightarrow\mw^C)-\kappa_{\mw}
\end{equation}
where $C_{wf}$ and $\kappa_{\mw}$ are as defined in Theorem \ref{thm:gauss}.
\end{corollary}
\begin{remark}
In \cite[Theorem 4.6]{aves:sub}, authors have shown that by quantization at noise level, Gaussian relay network achieves the cut-set bound within $14V$ bits. But Corollary \ref{cor:gauss} implies that quantization at noise level achieves the cut-set bound within $\dfrac{3}{2} V-1$ bits; thus we have tightened the gap between the achievable rate and the cut-set bound. A similar result holds for the two-way Gaussian relay network.
\end{remark} 
\section{conclusions}\label{sec:9}
We derived sufficient and necessary conditions for reliable multicasting of DMCS over cooperative networks. Necessary conditions were based on the cut-set type outer bound for the relay network. Sufficient conditions are based on joint source-channel coding, compress and forward strategy for the relay network and an identity related to the sub-modularity property of the entropy function. We showed that the sufficient conditions are indeed necessary conditions for some classes of deterministic networks including Aref networks and the linear finite-field deterministic networks. We also proved that multicasting of DMCS whose Slepian-Wolf region intersects the cut-set outer bound within a constant number of bits are feasible. In particular, we reduced all results of the paper to obtain achievable rate regions for multiple messages-multicast over the cooperative relay networks. We showed that this achievable rate region subsumes some recent achievable rate (region) for relay networks.      

\appendices
\section{Proof of Lemma \ref{le:cover}}\label{app:1}
We prove this lemma by contradiction. Let $d$ be the dimension of $\bp$. Suppose $\mf$ is not a closed covering of $\bp$, so there exists a point $A$ inside $\bp$ which is not covered by $\mf$ (Note that by assumption 2, the points that lie on the boundary of $\bp$ are covered). Let $B$ be the closest point in $\cup_{i=1}^n\bp_i$ to $A$. It is clear that $B$ must lie on a facet of at least one of the polytopes $(\bp_i:1\le i\le n)$. Denote this facet by $\F_{\bp_j}$. Two situations arise:
\begin{enumerate}
\item $\F_{\bp_j}$ lies inside $\bp$. Now by assumption 3, there exists $k\neq j$, such that $\bp_j\cap\bp_k=\F_{\bp_j}$. Let $\mathbf{S}(B,\epsilon)$ be a $d$-\emph{dimensional} sphere with center $B$ and radius $\epsilon$ which is small enough such that $\mathbf{S}(B,\epsilon)$ is contained in $\bp_j\cup\bp_k$. Then the segment $AB$ intersects $\mathbf{S}(B,\epsilon)$ at a point $C$ which belongs to one of $\bp_j$ or $\bp_k$. Now $C$ is closer than $B$ to $A$ and lies on $\cup_{i=1}^n\bp_i$. This results in contradiction, which proves lemma in this case.
\item $\F_{\bp_j}$ lies on the boundary of $\bp$. Let $\mathbf{S}(B,\epsilon)$ be a sphere with center $B$ and radius $\epsilon$ which is small enough such that $\mathbf{S}(B,\epsilon)$ only intersects $\bp_j$. Since, $A$ lies inside $\bp$, the segment $AB$ intersects $\mathbf{S}(B,\epsilon)$ at a point $C$ inside $\bp$. By assumption, $C$ belongs to $\bp_j$, which again results in contradiction that proves the lemma.      
\end{enumerate}    
\section{Proof of Lemma \ref{le:facet}}\label{app:2}
Denote the RHS of \eqref{eq:app} by $\F^*_{f,\mt}$. First, we prove that $\F^*_{f,\mt}\subseteq\F_{f,\mt}$. Suppose $\ux$ belongs to $\F^*_{f,\mt}$. Now for each $\mu\subseteq\mv$, we have:
\begin{align}
x_{\mu}&=x_{\mu\cap\mt}+x_{\mu\cap\mt^C}\label{eq:app21}\\
&\ge f(\mu\cap\mt|(\mu\cap\mt)^C)+f(\mu\cap\mt^C|\mt^C\cap\mu^C)\label{eq:app22}\\
&=f(\mv)-f(\mu^C\cup\mt^C)+f(\mt^C)-f(\mu^C\cap\mt^C)\label{eq:app23}\\
&\ge f(\mv)-f(\mu^C)\label{eq:app24}\\
&=f(\mu|\mu^C)\label{eq:app25}
\end{align}
where \eqref{eq:app22} follows from the definition of $\F^*_{f,\mt}$ and \eqref{eq:app24} follows, since $f$ is a sub-modular function. Now, \eqref{eq:app25} yields $\ux\in\F_{f,\mt}$. Hence $\F^*_{f,\mt}\subseteq\F_{f,\mt}$. On the other side, assume $\ux\in\F_{f,\mt}$. Note that by definition, $\ux_{\mt}\in\F^{(1)}_{f,\mt}$. For each $\ms\subseteq\mt^C$, consider:
\begin{align}
x_{\ms}&=x_{\mt\cup\ms}-x_{\mt}\\
&=x_{\mt\cup\ms}-f(\mt|\mt^C)\label{eq:app26}\\
&\ge f(\mt\cup\ms|\mt^C\cap\ms^C)-f(\mt|\mt^C)\\
&=f(\mt^C)-f(\mt^C\cap\ms^C)\\
&=f(\ms|\mt^C\backslash\ms)\label{eq:app27}
\end{align} 
where \eqref{eq:app26} follows, because $\ux$ lies on the hyperplane $x_{\mt}=f(\mt|\mt^C)$. Now, \eqref{eq:app27} implies that $\ux_{\mt^C}\in\F^{(2)}_{f,\mt}$ which results in $\F_{f,\mt}\subseteq\F^*_{f,\mt}$. Thus $\F^*_{f,\mt}=\F_{f,\mt}$. \par
Next, we show that $\F_{f,\mt}^{(1)}=\bp_{f_1}$ and $\F_{f,\mt}^{(2)}=\bp_{f_2}$. First observe that since $f$ is sub-modular, $f_1$ and $f_2$ are sub-modular functions. Hence $\bp_{f_1}$ and $\bp_{f_2}$ are well defined. Moreover, note that
\begin{align}
\forall\ms\subseteq\mt:
f_1(\ms|\mt\backslash\ms) &= f_1(\ms\cup\mt)-f_1(\mt\backslash\ms)\nonumber\\
                         &= f(\ms\cup\mt|\mt^C)-f(\mt\backslash\ms|\mt^C)\nonumber\\
                        &= f(\mv)-f([\mt\backslash\ms]\cup\mt^C)\nonumber\\
                       &= f(\mv)-f(\ms^C)\nonumber\\
                       &= f(\ms|\ms^C)\label{eq:eq}
\end{align}            
Comparing \eqref{eq:eq} and \eqref{eq:pface1} with Definition \ref{def:associate}, we conclude that $\F_{f,\mt}^{(1)}$ is the essential polytope of $f_1$ with dimension $|\mt|-1$. Likewise, we can show that $\F_{f,\mt}^{(2)}$ is the essential polytope of $f_2$ with dimension $|\mt^C|-1$.  This completes the proof.
\section{Formal Proof of Equation \eqref{eq:cl3}}\label{app:3}
By Lemma \ref{le:facet}, it suffices to prove the following identities:
\begin{align}
\ms\subseteq\mt^C:\quad & h_{\bC}(\ms|\mt^C\backslash\ms)=h_{\bC^*}(\ms|\ms^C)\\
\ms\subseteq\mt:\quad & h_{\bC}(\ms|\ms^C)=h_{\bC^*}(\ms|\mt\backslash\ms)
\end{align}
We prove the first identity. Proof of the second identity is similar. For each $\ms\subseteq\mt^C$ consider,
\begin{align}
h_{\bC}(\ms|\mt^C\backslash\ms)&=h_{\bC}(\mt^C)-h_{\bC}(\mt^C\cap\ms^C)\\
&=\sum_{k=1}^{K+1}H(X_{\mt^C\cap\ml_k}Y_{\mt^C\cap\ml_{k-1}}|X_{\ml^k}Y_{\ml^{k-1}})-H(X_{\mt^C\cap\ms^C\cap\ml_k}Y_{\mt^C\cap\ms^C\cap\ml_{k-1}}|X_{\ml^k}Y_{\ml^{k-1}})\\
&=\sum_{k=1}^{K+1}H(X_{\ms\cap\ml_k}Y_{\ms\cap\ml_{k-1}}|X_{\mt^C\cap\ms^C\cap\ml_k}Y_{\mt^C\cap\ms^C\cap\ml_{k-1}}X_{\ml^k}Y_{\ml^{k-1}})\label{eq:app:c1}
\end{align} 
Note that $\ml^{*k}=\ml^{k-1}\cup(\mt^C\cap\ml_{k-1})$. Moreover, for each $\ms\subseteq\mt^C$, simple calculations yield:
\begin{align}
\ms\cap\ml_k^*&=\ms\cap\left[(\mt\cap\ml_{k-1})\cup(\mt^C\cap\ml_k)\right]=\ms\cap\ml_k\nonumber\\
\ms^C\cap\ml_k^*&=\left[\mt\cap\ml_{k-1}\right]\cup\left[\ms^C\cap\mt^C\cap\ml_k\right]\nonumber\\
\ml^{*k}\cup(\ms^C\cap\ml_k^*)&=\ml^k\cup\left[\ms^C\cap\mt^C\cap\ml_k\right]\label{eq:app:c2}
\end{align}
substituting \eqref{eq:app:c2} in \eqref{eq:app:c1} gives:
\begin{align}
h_{\bC}(\ms|\mt^C\backslash\ms)&=\sum_{k=1}^{K+1}H(X_{\ms\cap\ml_k^*}Y_{\ms\cap\ml_{k-1}^*}|X_{\ms^C\cap\ml_k^*}Y_{\ms^C\cap\ml_{k-1}^*}X_{\ml^{*k}}Y_{\ml^{*k-1}}Z )\\
&=\sum_{k=1}^{K+2}H(X_{\ms\cap\ml_k^*}Y_{\ms\cap\ml_{k-1}^*}|X_{\ms^C\cap\ml_k^*}Y_{\ms^C\cap\ml_{k-1}^*}X_{\ml^{*k}}Y_{\ml^{*k-1}}Z)\\
&=h_{\bC^*}(\ms|\ms^C)
\end{align}
where in the last step, we have used the fact that $\ms\cap\ml_{K+1}^*=\ms\cap\ml_{K+1}=\emptyset$. This completes the proof.$\square$
\section{Equivalence of Constraints \eqref{eq:adi} and \eqref{eq:adi-adi}}\label{app:simplify}
It is sufficient to show that the RHS of \eqref{eq:adi1} and \eqref{eq:adi-adi1} are equal. Substituting $R_v=H(X_v)+H(\hY_v|X_vY_v)$ in the RHS of \eqref{eq:adi1} gives,
\begin{align}
R_{\mw}^{(d_i)}-H(\hY_{\mw}X_{\mw}|X_{\mw^C}\hY_{\mw^C\backslash\{d_i\}}Y_{d_i})&=H(X_{\mw})+H(\hY_{\mw}|X_{\mw}Y_{\mw})-H(\hY_{\mw}X_{\mw}|X_{\mw^C}\hY_{\mw^C\backslash\{d_i\}}Y_{d_i})\label{eq:sal:7}\\
&=I(X_{\mw};\hY_{\mw^C\backslash\{d_i\}}Y_{d_i}|X_{\mw^C})+H(\hY_{\mw}|X_{\mw}Y_{\mw})\n&\qquad\qquad -H(\hY_{\mw}|X_{\mv}\hY_{\mw^C\backslash\{d_i\}}Y_{d_i})\n
&=I(X_{\mw};\hY_{\mw^C\backslash\{d_i\}}Y_{d_i}|X_{\mw^C})-I(Y_{\mw};\hY_{\mw}|X_{\mv}\hY_{\mw^C\backslash\{d_i\}}Y_{d_i})\label{eq:sal:8}
\end{align}
where \eqref{eq:sal:7} follows from the fact that $d_i\notin\mw$, $X_t$'s are independent and $\hY_t$ given $(X_t,Y_t)$ is independent of all other random variables
 and \eqref{eq:sal:8} follows, since $(X_{\mw^C}\hY_{\mw^C}Y_{d_i})-(X_{\mw},Y_{\mw})-\hY_{\mw}$ forms a markov chain. Substituting \eqref{eq:sal:8} in \eqref{eq:adi1} shows that \eqref{eq:adi1} and \eqref{eq:adi-adi1} are equal. Also, using \eqref{eq:sal:8} with $\mw=\ms$ shows that \eqref{eq:adi2} and \eqref{eq:adi-adi2} are equal.   

 \section{Proof of Lemma \ref{le:9}}\label{app:5}
According to the codebook generation and the definition of $\mn_{\ms,\mz}(\uu_{\ma[b-\ell-V+2]},\cdots,\uu_{\ma[b-V+1]},\mathbf{s}_b)$  in \eqref{sal:def-mn}, $(\bX_t(s_t):t\in\ms\cap\ml_k)$ and $(\bX_{\mv}(s_{\mv,[b-k+1]}))$ are drawn independently from the sets $\stypt(X_t)$ and $\styp(X_{\mv})$. Also given $\bX_{t',[b-k+1]}( t'\in\mz\cap\ml_{k-1})$, $\bhY_{t'}(z_{t'}|\bX_{t',[b-k+1]})$ is drawn uniformly from the set $\styp(\hY_t|\bX_{t',[b-k+1]})$ and is independent from other random variables. Hence the joint p.m.f. of\\ $(\ux_{\ms\cap\ml_k}(s_{\ms\cap\ml_k}),\ux_{\mv}(s_{\mv,[b-k+1]}),\uhy_{\mz\cap\ml_{k-1}}(z_{\mz\cap\ml_{k-1}}),\uhy_{\ml^{k-1}\cup(\ml_{k-1}\backslash\mz),[b-k+1]},\uy_{d_i,[b-k+1]})$ factors as 
\be\label{sal:p}
\mathbb{P}[\ux_{\mv}(s_{\mv,[b-k+1]}),\uhy_{\ml^{k-1}\cup(\ml_{k-1}\backslash\mz),[b-k+1]},\uy_{d_i,[b-k+1]}]\prod_{t\in\ms\cap\ml_k}P_{\bX_t}(\ux_{t}(s_t))\prod_{t'\in\mz\cap\ml_{k-1}}P_{\bhY_t|\bX_t}(\uhy_{t'}(z_{t'}|\ux_{t',[b-k+1]})),
\ee
where $P_{\bX_t}$ and $P_{\bhY_t|\bX_t}$ are uniform distributions on the sets $\stypt(X_t)$ and $\stypo(\hY_t|X_t)$, respectively. Now, we upper bound $\mathbb{P}[\me_3(b,k,\mathbf{s})]$ for each $\mathbf{s}\in \mn_{\ms,\mz}(\uu_{\ma[b-\ell-V+2]},\cdots,\uu_{\ma[b-V+1]},\mathbf{s}_b)$ as follows,
\begin{align}
\mathbb{P}[\me_3(b,k,\mathbf{s})]&=\sum_{\mathrlap{\big(\ux_{\ml_k\backslash\ms}(s_{\ml_k\backslash\ms,[b-k+1]}),\uhy_{\ml_{k-1}\backslash\mz}(z_{\ml_{k-1}\backslash\mz,[b-k]}),\ux_{\ml^k,[b-k+1]},\uhy_{\ml^{k-1},[b-k+1]},\ux_{d_i},\uy_{d_i}\big)\in\styp}}
\mathbb{P}[\ux_{\ml_k\backslash\ms}(s_{\ml_k\backslash\ms,[b-k+1]}),\uhy_{\ml_{k-1}\backslash\mz}(z_{\ml_{k-1}\backslash\mz,[b-k]}),\uhy_{\ml^{k-1}[b-k+1]},\ux_{d_i,[b-k+1]},\uy_{d_i,[b-k+1]}]\n
&\qquad\quad\sum_{\mathclap{\ux_{\ms\cap\ml_k}(s_{\ms\cap\ml_k}),\uhy_{\mz\cap\ml_{k-1}}(z_{\mz\cap\ml_{k-1}})\in\atop\styp(X_{\ms\cap\ml_k}\hY_{\mz\cap\ml_{k-1}}|\ux_{\ml_k\backslash\ms,[b-k+1]},\uhy_{\ml_{k-1}\backslash\mz,[b-k]},\uhy_{\ml^{k-1}[b-k+1]},\uy_{d_i,[b-k+1]})}}\qquad\qquad\qquad\prod_{t\in\ms\cap\ml_k}P_{\bX_t}(\ux_{t}(s_t))\prod_{t'\in\mz\cap\ml_{k-1}}P_{\bhY_t|\bX_t}(\uhy_{t'}(z_{t'}|\ux_{t',[b-k+1]}))\label{sal:p1}\\
&=\mathbb{P}[(\bX_{\ml_k\backslash\ms}(s_{\ml_k\backslash\ms,[b-k+1]}),\bhY_{\ml_{k-1}\backslash\mz}(z_{\ml_{k-1}\backslash\mz,[b-k]}),\bhY_{\ml^{k-1}[b-k+1]},\bX_{d_i,[b-k+1]},\bY_{d_i,[b-k+1]})\in\styp]\n&
\qquad\dfrac{|\styp(X_{\ms\cap\mz}\hY_{\mz\cap\ml_{k-1}}|X_{\ml_k\backslash\ms}\hY_{\ml_{k-1}\backslash\mz}X_{\ml^{k}}\hY_{\ml^{k-1}}X_{d_i}Y_{d_i})|}{\prod_{t\in\ms\cap\ml_k}|\stypt(X_t)|\prod_{t'\in\mz\cap\ml_{k-1}}|\stypo(\hY_{t'}|X_{t'})|}\label{sal:p2}\\
&\stackrel{.}{\le} 2^{-n\beta_{\ms,\mz}(k)}\label{sal:p3}
\end{align}
where \eqref{sal:p1} follows from \eqref{sal:p}, \eqref{sal:p2} follows from the definition of $P_{X_t}$ and $P_{\hY_t|X_t}$ and \eqref{sal:p3} is a result of the properties of jointly typical sequences. 
\section{Proof of Lemma \ref{le:7}}\label{app:6}
According to the definition of $\mn_{\ms,\mz}(\uu_{\ma[b-\ell-V+2]}\cdots,\uu_{\ma[b-V+1]},\mathbf{s}_b)$, for $\mathbf{s}=(w_{\ml_1},z_{\ml_1},\cdots,w_{\ml_{\ell}},z_{\ml_{\ell}})$ in \eqref{sal:def-mn}, each $(z_v:v\in\mz)$ takes $2^{n(I(\hY_v;Y_v|X_v)+\delta)}-1$ different values and each $(z_v:v\in\mv_{-d_i}\backslash\mz)$ takes a fixed value, thus $z_{\mv_{-d_i}}$ takes less than $2^{n(\sum_{t\in\mz}I(Y_t;\hY_t|X_t))}$ different values. Also, according to the definition for each $k\in[1,\ell]$, $w_{\ml_k\backslash\ms}$ takes the fixed value $w_{\ml_k\backslash\ms,[b-k+1]}$ and $w_{\ms\cap\ml_k}$ must satisfy the following relation:
\[
\uu_{\ms\cap\ml_k}(w_{\ml_k\cap\ms})\in\styp(U_{\ml_k}(w_{\ml_k\cap\ms})|\uu_{\ml_k\backslash\ms}(w_{\ml_k\backslash\ms,[b-k+1]}),\uu_{\ml^k,[b-k-V+2]},\uu_{d_i,[b-k-V+2]}).
\]  
Thus $\uu_{\ms\cap\ml_k}(w_{\ml_k\cap\ms})$ (or equivalently $w_{\ml_k\cap\ms}$) takes at most $2^{n(H(U_{\ms\cap\ml_k}|U_{\ml_k\backslash\ms}U_{\ml^k}U_{d_i}))}$ different values. Therefore, $w_{\mv_{-d_i}}$ takes at most $2^{n(\sum_{k=1}^{\ell}H(U_{\ms\cap\ml_k}|U_{\ml_k\backslash\ms}U_{\ml^k}U_{d_i}))}$ different values. Now, comparing the bounds on the number of possible choices for $z_{\mv_{-d_i}}$ and $w_{\mv_{-d_i}}$ yields the lemma.
\section*{acknowledgement}
We would like to thank the anonymous reviewers and the Associate Editor for their suggestions which greatly improved the paper in terms of its presentation as well as technical clarity and context. We would also like to thank the members of Information Theory and Security Lab at Sharif University for their comments.
 
\begin{IEEEbiographynophoto}{Mohammad Hossein Yassaee} 
received the B. S. and M. S. degrees in electrical engineering from Sharif University of Technology, Tehran, Iran, in 2007 and 2009, respectively.\par
He is currently pursuing the Ph.D. degree at the Sharif University of Technology, Tehran, Iran under the supervision of Prof. M. R. Aref. His research interests are in the areas of information theory and probability. These include relay networks and information theoretic security.  
\end{IEEEbiographynophoto}
\begin{IEEEbiographynophoto}{Mohammad Reza Aref} 
 was born in city of Yazd in Iran in 1951. He received his B.S.
in 1975 from University of Tehran, his M.S. and Ph.D. in 1976 and 1980, respectively,
from Stanford University, all in Electrical Engineering. He returned to Iran in 1980 and
was actively engaged in academic and political affairs. He was a Faculty member of
Isfahan University of Technology from 1982 to 1995. He has been a Professor of Electrical
Engineering at Sharif University of Technology since 1995 and has published more than 190
technical papers in communication and information theory and cryptography in international
journals and conferences proceedings. His current research interests include areas of
communication theory, information theory and cryptography with special emphasis on
network information theory and security for multiuser wireless communications. At the same
time, during his academic activities, he has been involved in different political positions.
First Vice President of I. R. Iran, Vice President of I. R. Iran and Head of Management and
Planning Organization, Minister of ICT of I. R. Iran and Chancellor of University of Tehran,
are the most recent ones.
\end{IEEEbiographynophoto}
\end{document}